\documentclass[11pt,a4paper]{article}

\def\bi{\begin{itemize}}
\def\ei{\end{itemize}}
\def\bea{\begin{eqnarray}} 
\def\eea{\end{eqnarray}}
\def\be{\begin{equation}}
\def\ee{\end{equation}}
\def\line{\hbox to \hsize}    
\def\frac #1#2{{#1\over #2}}

\def\tr{{\rm  tr\,}}

\def\brak #1#2{{\langle#1, #2\rangle}}

\def\1{\mbox{\bf I}}

\def\sech{{\rm sech\,}}
\def\tanh{{\rm tanh\,}}
\def\csch{{\rm csch\,}}
\def\coth{{\rm coth\,}}
\def\csch{{\rm csch\,}}
\def\cosec{{\rm cosec\,}}

\newenvironment{Quote}
{\begin{list}{}{%
\setlength{\leftmargin}{10 pt}
\setlength{\rightmargin}{\leftmargin}}
\item[]}
{\end{list}}

     {\refstepcounter{exercisenumber} \begin{Quote}\small{\sl
     Exercise~\thechapter.\theexercisenumber\/}:}
{\end{Quote}}

{\begin{Quote}\small{\sl Example\/}:}
{\end{Quote}}

\def\levelonelist{
        \begin{list}{\mybulA}%
                        {
        \setlength{\topsep}{0pt}
        \setlength{\parsep}{0pt}
        \setlength{\partopsep}{0pt}
        \setlength{\itemsep}{0pt}
                        }
                }
\def\leveltwolist{
        \begin{list}{\mybulB}%
                        {
        \setlength{\topsep}{0pt}
        \setlength{\parsep}{0pt}
        \setlength{\partopsep}{0pt}
        \setlength{\itemsep}{0pt}
                        }
                }
\def\bo{\levelonelist}

\def\el{\end{list}}

\usepackage{jheppub}
\usepackage{url}
\usepackage{amsfonts}
\usepackage{amssymb}

\input epsf

\usepackage{color}   
\usepackage{hyperref} 
\hypersetup{
    linktocpage={true}
    colorlinks=true, 
    linktoc=all,     
    linkcolor=blue,  
}

\begin{document}
\title{{\bf Radon transforms and the SYK model}}
\author{Michael Stone}
\affiliation{University of Illinois at Urbana-Champaign\\Loomis Laboratory of Physics\\
                                           1110 W Green Street\\ Urbana, IL}

\abstract{Motivated by recent work on the Sachdev-Ye-Kitaev (SYK) model, we consider the effect of Radon or X-ray transformations, on the Laplace eigenfunctions in hyperbolic Bolyai-Lobachevsky space. We show that the Radon map from this space to Lorentzian-signature  de Sitter space is easier to interpret if we use the Poincare disc model and eigenfunctions rather than the upper-half-plane model. In   particular, this version of the  transform   reveals the geometric origin  of the boundary conditions imposed on the eigenfunctions that are involved in  calculating  the SYK four-point function.}                                           
\maketitle

\vfil\eject

\section{Introduction}

The  Sachdev-Ye-Kitaev (SYK) model \cite{kitaev-syk,sy,maldacena} is a one-dimensional quantum-mechanics  model with interesting physical properties. It is 
solvable at strong coupling, maximally chaotic, and  has an  emergent conformal symmetry. There are  strong hints that it has a AdS/CFT gravity dual. For reviews see \cite{rosenhaus,sarosi}.

The present paper is not so much about the physics of this model, but rather is about some mathematics  that appears   in its analysis.  In solving the bi-local integral equation that arises when computing the 4-point correlation function, the authors of \cite{joeP} need to find the  eigenfunctions of a Schr{\"o}dinger operator  whose potential is unbounded below. Although this seems an ill-defined problem, they were   able to find  a complete  orthonormal set of suitable eigenfunctions by imposing some   boundary condition infinitely deep in the potential abyss.  The resulting spectrum has  both a continuous set of scattering states and a discrete set of normalizable bound states --- both sets of eigenfunctions involving   an unusual linear combination Bessel functions. 

It turns out that the     unbounded-below  Schr{\"o}dinger  operator   has been previously discussed  in the literature \cite{dunster90,gitman,andrianov}, and it is known that it possesses  a one-(or even two-\cite{kobayashi})-parameter family of  boundary conditions each of which  lead to a self-adjoint operator and hence   a complete set of eigenfunctions. The particular   boundary condition selected from this infinite set by the authors of \cite{joeP}  was required     to  ensure compatibility  with the   ${\rm SL}(2, {\mathbb R})$ symmetry of the integral equation. The analytic necessity is clear, but given that the SYK model is expected  to describe some sort of quantum geometry it would be useful to find a   geometric principle   behind  the selection.  

A key hint comes from the    observation   \cite{jevicki}  that the  peculiar combination of Bessel functions that compose the selected  eigenfunctions also arises  as the result of the application of  a    Radon, or X-ray, transformation  to   the modified Bessel-function eigenfunctions  of the Laplacian on the Poincar{\'e} upper-half-plane $H_+^2$.   

There is an extensive literature on  Radon transforms on hyperbolic spaces \cite{helgason,gelfand,hyperbolic_radon} but we   seldom find in it   explicit transforms of individual   $L^2[H_+^2]$ eigenfunctions.   
The authors of  \cite{jevicki} evaluate the Radon transform integral by using a Bessel  identity from Watson's Treatise  \cite{watson} (\S 12.11, Eq.4). There are very many Bessel-function identities, but they tend to fall into families that either reflect general properties of hypergeometric functions or have some geometric origin. This one seems  rather obscure and not   associated with such a family. The  purpose of the present  paper is explore how this and some other ``special function''  identities   relate to  the geometry of the  Radon transformation. In particular we will see that evaluating  the Radon transformation of  the eigenfunctions of the  Laplacian in  the Poincar{\'e} disc model (as opposed to those of the upper-half-plane model)  reveals   why the particular choice of self-adjoint extension is special.

In order to have the necessary geometry in our mind, in the next section we  review the geometry of hyperbolic space and its  two standard models: the Poincar\'e disc and the upper half plane. We then do the same for  the Lorentzian-signature analogue AdS/dS. We show how  actions of the isometry  groups   $SL(2,{\mathbb R})$ and ${\rm SU}(1,1)$  on the space of  geodesics leads to their parametrization as cosets, and in section \ref{SEC:eigen} we obtain the normalized   eigenfunctions  that provide the spectral decompositions on these cosets. In section \ref{SEC:radon} we use the asymptotic properties of the eigenfunctions to evaluate their Radon transform integrals. In the case of upper-half-plane we simply  recover  the results of \cite{jevicki}, but for the Poincar\'e disc  the transform  leads to some rarely-met  special functions and  at the same time reveals the origin and necessity of the boundary conditions   selected  in \cite{joeP}. Two appendices review how the coset metrics arise from ${\rm SL}(2,{\mathbb R})\simeq {\rm SU}(1,1)$, and derive the   orthogonality properties of the rarely-met special functions.

\section{Geometry}
\label{SEC:geometry}

\subsection{Hyperbolic  Space}
\label{SUBSEC:hyperbolic}

Hyperbolic or Bolyai-Lobachevsky space 
 arises  from  the embedding  the upper sheet of the two-sheeted hyperboloid 
\be
T^2-X^2-Y^2=1
\label{EQ:hyperboloid}
\ee
into a  Minkowski space with metric 
\be
ds^2= -dT^2+dX^2+dY^2.
\ee
Despite the minus sign in the Minkowski metric,  the resulting Riemann manifold  has a positive definite   metric with  constant negative gaussian curvature $\kappa=- 1$. The   Ricci-scalar curvature  is  $R=-2$.

\begin{figure}[tbp]
\centering 
\includegraphics[width=.45\textwidth]{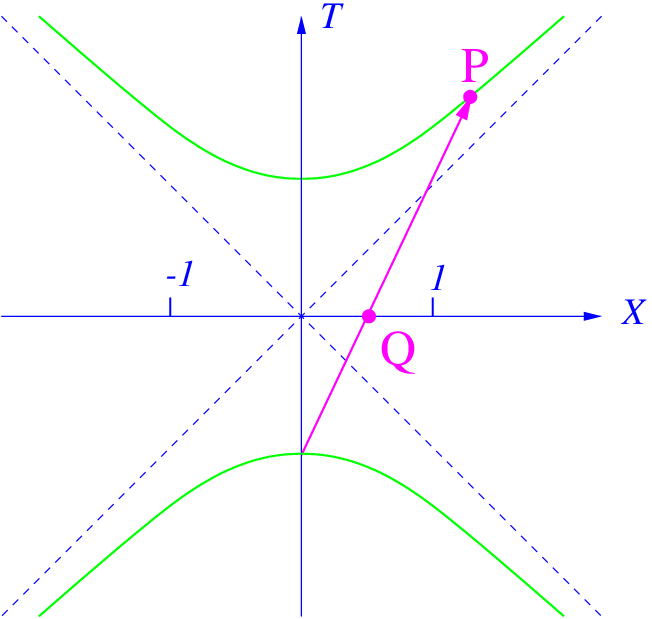}
\caption{\label{FIG:1} The figure shows  the $Y=0$ slice through a hyperbolic analogue of the familiar  $S^2 \to {\mathbb C}$ stereographic map.  The map   takes a point ${\rm P}=(X,Y,T)$ on the upper sheet of the hyperboloid to ${\rm Q}=(X,Y)$  in  the $T=0$ plane. As $T\to \infty$ for $X>0$   and  $Y=0$  the point  $Q$ approaches $(0,0,1)$ and so Q always lies in  the  disc   $X^2+X^2<1$.}
\end{figure}

We  parametrize the  hyperboloid (\ref{EQ:hyperboloid})
by 
\be
T=  \cosh \rho,\quad X+iY= e^{i\phi} \sinh \rho 
\ee
and 
recall the hyperbolic ``$t$'' substitution $t= \tanh(\rho/2)$, whence
\be
\cosh \rho= \frac{1+t^2}{1-t^2},\quad \sinh \rho = \frac{2t}{1-t^2}.
\ee
Figure \ref{FIG:1}  shows that the tangent of the angle $\theta$  between PQ and the $T$ axis is
\be
\tan \theta  = \frac { \sinh \rho}{1+ \cosh \rho}=\tanh (\rho/2),
\ee
 and so  we have a  explicit formula 
\be
Z= X+iY = e^{i\phi} \, \tanh (\rho/2)=  e^{i\phi}t
\ee
for the coordinates of Q in terms of those of P.

In the  hyperbolic analogue  of spherical polar coordinates we assign ${\rm P}\to (\rho,\phi)$ and  the metric is 
\be
ds^2 =d\rho^2 + (\sinh \rho)^2 d\phi^2.
\ee
Parametrizing the point P on the hyperboloid  by    the  $Q=(X,Y)$ coordinates in  the $T=0$ plane, the metric becomes 
\be
ds^2 = \frac{4 }{(R^2-X^2-Y^2)^2} (dX^2+dY^2), \quad X^2+Y^2<1.
\ee
This is  the  {\it Poincar{\'e} disc\/} model $D^2_P$ of Lobachevsky space.

A third parametrization of Lobachevsky space is the {\it upper half-plane} $H^2_+= \{(x,y)\in {\mathbb R}^2\,|\,y>0\}$. If we set $z=x+iy$ and $Z=X+iY$,  the upper half-plane is  mapped 1-1 and conformally onto the Poincar\'e disc   by 
\be
Z=\frac{z-i}{z+i}\quad  \Leftrightarrow \quad z= i \frac{1+Z}{1-Z}.
\ee
The Lobachevsky metric in upper-half-plane coordinates becomes  
\be
ds^2 = \frac 1{y^2} (dx^2+dy^2).
\ee
If  points on the  boundary of the Poincar{\'e} disc  are  parameterized by $Z= -e^{i\theta}$  then $x= \tan(\theta/2)$ is the corresponding point on the $y=0$ boundary of the upper half-plane.

\subsection{Lorentzian signature: AdS$_2$ and dS$_2$}
\label{SUBSEC:AdS}


 Hyperbolic Bolyai-Lobachevsky space is a  Euclidean-signature version of two dimensional Anti-de Sitter space AdS$_2$.   
The Lorentz-signature AdS$_2$ is obtained by the embedding of the \underline{one}-sheeted hyperboloid 
\be
u^2+v^2-r^2=1
\ee
into a space with metric
\be
ds^2 =-du^2-dv^2+dr^2.
\ee


We can exploit   the identity
\be
{\sec^2}\sigma- {\tan^2}\sigma=1
\ee to 
parametrize the  hyperboloid
by
\bea
r&=& \tan\sigma,\quad  -\pi/2<\sigma<\pi/2\nonumber\\
u&=& \cos \tau \sec \sigma, \nonumber\\
v&=& \sin \tau \sec \sigma.
\eea
The resulting  coordinate chart covers the entirety of the AdS$_2$ hyperboloid. The coordinate    $\tau$ is periodic  $\tau\sim \tau+2\pi$. In AdS$_2$     $\tau$  is  the time coordinate  so  there are  closed time-like loops. Such  loops are usually undesirable and it is customary to extend to a simply-connected  covering space. We will not make this extension  in the present paper because 
interchanging the role of the time coordinate $\tau$ and space coordinate $\sigma$ turns anti-de Sitter space  into ordinary de Sitter space dS$_2$ --- a  space-like circle whose circumference is exponentially increasing in time --- and our  Radon transformations  are  more naturally regarded  as  maps into de Sitter space with its periodic spatial coordinate. 

As  $d (\tan \sigma) = {\sec^2}\sigma\,d\sigma$, $d( \sec\sigma)= \sec\sigma \tan\sigma\, d\sigma$, this parametrization   induces the  metric 
\bea
ds^2 &=& (- {\sec^2} \sigma\, d\tau^2 - {\sec^2} \sigma \,{\tan^2} \sigma \, d \sigma^2) + {\sec^4}\sigma d\sigma^2\nonumber\\ 
&=& {\sec^2} \sigma (-d\tau^2+d\sigma^2).
\label{EQ:AdSGlobal1}
\eea

A different global   coordinate system is  given by exploiting the identity ${\cosh^2}\rho-{\sinh^2}\rho =1$ to  parametrize the hyperboloid as 
 \bea
 r&=& \sinh \rho,\nonumber\\
 u&=& \cosh \rho \cos \tau,\nonumber\\
 v&=& \cosh \rho \sin  \tau.
 \eea
 Now we have  
\bea
ds^2 &=& -{\cosh^2} \rho\, d\tau^2+d\rho^2,\nonumber\\
&=& -(1+r^2) d\tau^2+ \frac{dr^2}{1+r^2}.
\eea

\begin{figure}
\centering
\includegraphics[width=.45\textwidth]{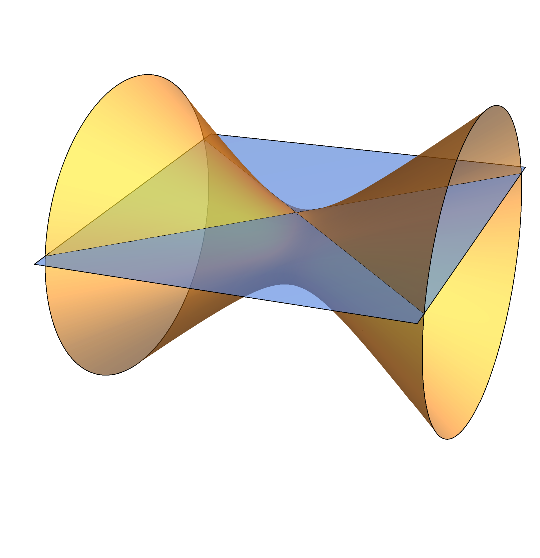}
\caption {\label{FIG:rindler} The intersection of the plane and hyperboloid. The Rindler $t$, $\varrho$ patch covers the  part of the hyperboloid above the  blue plane.} 
\end{figure}

Both the previous coordinate systems cover the whole hyperboloid. A coordinate chart  that covers only {\it part\/} of the space is given by parameterizing
\bea
 u&=& \cosh \varrho,\nonumber\\
 r+v&=& e^t  \sinh \varrho, \nonumber\\
 r-v &=& e^{-t}  \sinh  \varrho, 
 \eea
 so that 
 \be
 1={\cosh^2}\varrho-{\sinh^2}\varrho=u^2-(r+v)(r-v) = u^2+v^2-r^2.
 \ee
 Equivalently 
\bea
 u&=& \cosh \varrho,\nonumber\\
 r&=& \cosh t  \sinh \varrho, \nonumber\\
 v &=& \sinh t \sinh  \varrho.
 \eea
 This chart  only covers the part of the hyperboloid with $u>1$.  The intersection of the $u=1$ surface with the hyperboloid is a pair of intersecting straight lines in the three-dimensional embedding space,  and the  coordinate patch is the part above the blue plane in figure \ref{FIG:rindler}.
 The intersecting straight lines are null geodesics. A pair of such  lines can be constructed at any point as the hyperboloid is a {doubly-ruled surface}.

The metric in this chart  is  
\be
ds^2 =- {\sinh^2} \varrho \,dt^2 + d\varrho^2.
\ee
This appears to be a simple replacement  $t\to it$ in the hyperbolic polar coordinate metric $ds^2=  {\sinh^2} \rho\, dt^2 + d\rho^2$, but the fact that the $\varrho,t$  coordinates only cover part of the manifold shows that the Euclidean $\leftrightarrow$ Lorentzian correspondence is not so simple. 

 As ${\sinh^2} \varrho= {\cosh^2} \varrho-1$ and $u= \cosh \varrho$ we  can also write
\be
ds^2= -  (u^2-1) dt^2 + \frac{du^2}{u^2-1},
\ee
which makes $u=1$ look like a horizon.
 It bears  the same relation to the hyperbolic polar  coordinates as the Lorentzian  Rindler wedge has to Euclidean plane polar coordinates. The projections onto the $u=1$ plane of the constant $\rho$ curves are hyperbol{\ae}, and the $t=$constant curves are straight lines through $r=v=0$ point.

\begin{figure}
\centering
\includegraphics[width=.45\textwidth]{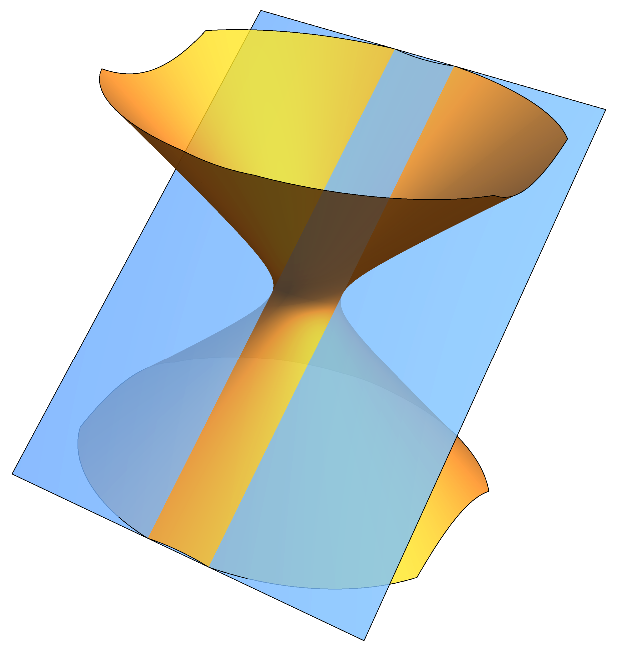}
\caption{\label{FIG:poincare} The two AdS$_2$ Poincar\'e coordinate patches  are divided  by  the plane $u-r=0$ which is shown in blue.} 
\end{figure}

Yet another parametrization  sets
\bea
u&=& \frac 1 {2\xi}(1+\xi^2-\xi^2\eta^2),\nonumber\\
v&=& \xi t,\nonumber\\
r&=& \frac 1 {2\xi} (1- \xi^2-\xi^2\eta^2).
\eea
We again have
\be
u^2+v^2 -r^2 =1
\ee
and now.
\be
ds^2 =- \xi^2\, d\eta^2+ \frac{d\xi^2}{\xi^2}.
\ee
If we set $\zeta = 1/\xi$, this becomes 
\bea
u&=& \frac 1{2\zeta}(\zeta^2+1-\eta^2)),\nonumber\\
v&=& \eta/\zeta,\nonumber\\
r&=& \frac1 {2\zeta} (\zeta ^2- 1-\eta^2)).
\eea
and
\be 
ds^2 =\frac{- d\eta^2+ d\zeta^2}{\zeta^2},
\ee
which is the Lorentz-signature version of the Poincar\'e upper-half-plane metric. The point $\zeta=0$ is not part of the chart. Because   
\be
 \frac 1 \zeta= \xi = u-r 
\ee
the plane $u-r=0$ divides the hyperboloid into two halves with $\xi \lessgtr 0$ (see figure \ref{FIG:poincare}).  One half  includes almost the entire boundary at  $ r\to+ \infty$ and only a small part  at  $r\to -\infty$, and {\it vice versa}. Note that when $u=r$ we have $v=\pm 1$, so the sides of  the orange  ``tail" are indeed parallel.

\begin{figure}
\centering
\includegraphics[width=.45\textwidth]{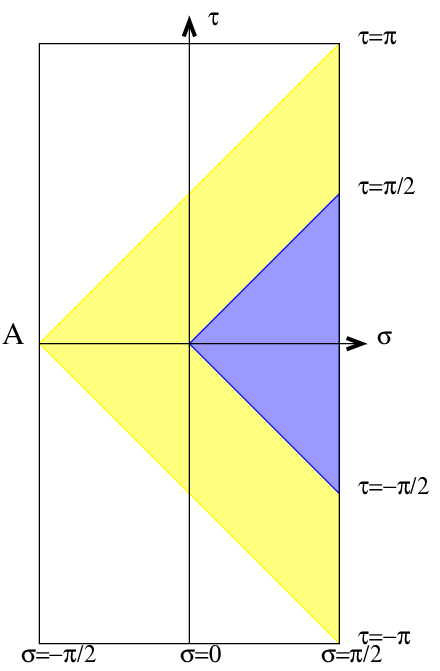}
\caption{\label{FIG:AdS}  AdS/dS coordinate patches: The rectangle is the $-\pi/2<\sigma<\pi/2$, $-\pi <\tau< \pi$ global chart. The blue region is the Rindler region covered by the $\varrho$, $t$ coordinates, and the yellow is a  $\zeta,\eta$ Poincar\'e chart. The part of the yellow region  with $\sigma<0$ is the parallel-sided orange ``tail" from figure \ref{FIG:poincare}. The point {\rm A}  only appears to be a single point because of the infinite scaling near $\sigma=- \pi/2$.  } 
\end{figure}

We can directly connect the Poincare chart  with the global chart by defining 
\be
\tan (u_\pm/2)= \eta \pm \zeta, 
\ee
with $\pi >u_+\ge u_- >-\pi$ so as to keep $\zeta>0$.
Then
\bea
ds^2 &=&\frac{ -d\eta^2+ d\zeta^2}{\zeta^2} \nonumber\\
&=& \frac {d(\zeta+\eta)d(\zeta-\eta)}{\zeta^2} \nonumber\\
&=& -4\frac{d[\tan (u_+/2)]d [\tan( u_-/2)]}{(\tan (u_+/2)-\tan (u_-/2))^2}
\eea
Using $d \tan x= {\rm sec}^2 x\,dx $ and $\sin(x-y)= \sin x \cos y-\cos x\sin y$ this becomes
\be
ds^2 = -\frac{d u_+ d u_-}{\sin^2\left(\frac{u_+-u_-}{2}\right)},
\ee
which is a slight variation on the first of our global AdS metric (\ref{EQ:AdSGlobal1}).  We will later have reason to relabel
\be
u_+\to  t_1, \quad u_-\to t_2, \qquad {\textstyle \frac 12} (t_1-t_2)=\alpha, \quad {\textstyle \frac 12} (t_1+t_2)=\theta,
\label{EQ:modifiedGlobal1}
\ee 
and then then  (\ref{EQ:modifiedGlobal1}) becomes
\be
ds^2= \frac{-d\theta^2+d\alpha^2}{\sin^2(\alpha)}, 
\ee
which   reappears as the metric (\ref{EQ:alpha-theta-metric}) and (\ref{EQ:alpha-theta-metric2}) for the space of geodesics on the Poincar\'e disc parametrized as in figure \ref{FIG:geodesicmap}.

\section{Isometries and Geodesics}
\label{SEC:geodesics}

Just as the positive-curvature  two-sphere has as  symmetry  the rotation group ${\rm SO}(3)$ which  acts {via\/}  its double cover ${\rm SU}(2)$, the negative-curvature  Bolyai-Lobachevski space  has  symmetry  group ${\rm SO}^+(2,1)$ which  acts {via\/}   its  double cover ${\rm SL}(2,{\mathbb R})$. The natural way of expressing  the action depends on the choice of coordinate system. 

\subsection{Isometries of    $H_+^2$}
 The M\"obius map
\be
z\mapsto z'= \frac{az+b}{cz+d}, \quad g=\left(\begin{matrix}a&b\cr c&d\end{matrix}\right) \in {\rm SL}(2, {\mathbb R}),
\ee
maps the upper half-plane  1-1 onto itself and preserves the metric 
\be
ds^2= \frac{dzd\bar z}{y^2}= \frac{dx^2+dy^2}{y^2}. 
\ee
  The M\"obius map is insensitive to the overall sign of the ${\rm SL}(2, {\mathbb R})$ matrix, so the group that is acting is really 
\be
{\rm PSL}(2, {\mathbb R}) \stackrel{\rm def}{=} {\rm SL}(2, {\mathbb R})/\{1,-1\}.
\ee 
This group   provides a continuous family of   orientation preserving  isometries  of $H^2_+$.  As the isometry group of the Minkowski-embedded hyperboloid  is the time-orientation preserving Lorentz group ${\rm SO}^+(2,1)$ we deduce that there is   2-to-1 mapping between ${\rm SL}(2, {\mathbb R})$ and ${\rm SO}^+(2,1)$. 

Elements of ${\rm SL}(2,{\mathbb R})$ are classified as {\it elliptic} if they fix a point in the interior of $H^2_+$, {\it parabolic\/}  if they possess  a single fixed point  on the boundary, and {\it hyperbolic}  if they fix \underline{two} points on the boundary. These conditions are determined by the trace: elliptic, parabolic, and hyperbolic elements have ${\rm tr}(g)$  respectively less than, equal to, or greater than two. 

The infinitesimal M\"obius maps
\be
\delta z= \frac{(1+\delta a)z +\delta b}{\delta c\, z+ (1+\delta d)} -z = (\delta b) + (\delta a - \delta d)z-(\delta c) z^2,\quad \delta a+\delta d=0,
\ee
generate the Lie algebra $\mathfrak{sl}(2, {\mathbb R})$. 
 This algebra  has a matrix representation  \cite{kitaev}
\be
L_{-1}=\left[\begin{matrix}0&1\cr 0 &0\end{matrix}\right], \quad L_{+1}= \left[\begin{matrix}0&0\cr -1 &0\end{matrix}\right], \quad L_0=\frac 12 \left[\begin{matrix}1&0\cr 0 &-1\end{matrix}\right],
\ee
with 
\be
[L_n, L_m]= (n-m)L_{n+m}. 
\ee
It is  convenient to select the sign for the  quadratic Casimir so that
\be
C_2\equiv-L_0^2+(L_{-1}L_{1}+ L_{-1}L_1)/2.
\ee
In the matrix representation  
\be
C_2\to  -\frac 34 {\mathbb I}.
\ee
 Using the $L_n$  matrices  we can write 
\be
\left(\begin{matrix} \delta a &\delta b\cr \delta c &\delta d\end{matrix}\right) = \delta b L_{-1} + (\delta a - \delta d)L_0- \delta c L_{+1}.
\ee
The  infinitesimal displacements $\delta z=\delta x+i \delta y $ define 
vector fields $
\delta x\, \partial_x+\delta y\, \partial_y
$
which   are  differential  operators 
\bea
L_{-1}&\mapsto& \frac{\partial}{\partial x}\nonumber\\
L_{+1}&\mapsto&(x^2-y^2) \frac{\partial}{\partial x}+2xy \frac{\partial}{\partial y}\nonumber\\
L_0&\mapsto& x \frac{\partial}{\partial x}+ y  \frac{\partial}{\partial y}\
\eea
 that   act on functions on  $H_+^2$.
The  Casimir operator in the differential operator  representation becomes    
\be
 C_2 \equiv -L_0^2 +{\textstyle\frac 12}(L_+L_-+L_-L_+)\mapsto  -y^2 \left( \frac{\partial^2}{\partial x^2}+ \frac{\partial^2}{\partial y^2}\right)
\ee
and coincides with  (minus) the Laplacian  on $H^2_+$.

With  $x=\xi$, $y=1+\eta$ with small $(\xi,\eta)$ we have 
\bea
\frac{L_{-1}+L_{+1}}{2}&=&  -\eta \frac{\partial}{\partial x} + \xi \frac{\partial}{\partial y}+ O(\xi^2,\eta^2, \xi\eta)\,
\eea
which describes  a rotation about $z=i$.  Thus the generator $(L_{-1}+L_{+1})/{2}$ exponentiates to a compact ${\rm SO}(2)$  elliptic subgroup $K$ of ${\rm SL}(2{\mathbb R})$ corresponding to the rotation in  the two-to-one ${\rm SL}(2,{\mathbb R})\simeq {\rm SO}^+(2,1)$ morphism.  

\subsection{Upper-half-plane geodesics} 
\label{SEC:half-plane-geodesics}

The isometries act transitively on 
the upper half-plane  geodesics, which  are semicircles with centres on the real axis.
Every  geodesic lies on  the isometry orbit of  a reference geodesic which has its centre at $x=0$ and with ends on the real axis at $x=\pm 1$.
  
To see how this works consider:
\bo
\item[$\bullet$] the hyperbolic subgroup  $A= \{ \exp \sigma(L_{-1}- L_1)/2\}$ leaves the points $x=\pm 1$ on the $x$-axis fixed and so preserves   the reference  geodesic as a set, although increasing $\sigma$  flows  points along the geodesic  from $x=-1$ towards $x=+1$.

\item[$\bullet$]the hyperbolic subgroup $H= \{\exp \xi L_0\} $ leaves fixed the points $x=0$ and $z=\infty$ and acts  on the reference geodesic to dilate its (Euclidean) radius from unity to $e^\xi$.

\item[$\bullet$]the parabolic subgroupgroup $\tilde N= \{\exp t L_{-1}\}$ leaves only the point at infinity fixed and rigidly translates  the geodesic semicircle along the $x$ axis through a distance $t$. 

\el
The last two operations generate can generate any upper-half-plane geodesic. Thus the space of   geodesics may  be identified with the coset ${\rm SL}(2,{\mathbb R})/A$  parameterized by $t,\xi$. 

The orbit of the point $z=i$ under the action of  
\be
g= \exp\{ t L_{-1}\} \exp \{\xi L_0\}\exp \{\sigma(L_{-1}- L_1)/2\}
\ee
is 
\bea
x(\sigma)&=&e^{\xi}\, \tanh \sigma+t\nonumber\\
y(\sigma)&=& e^{\xi}\, \sech\sigma
\eea
and so $\sigma$ parameterizes points on the particular geodesic $\gamma(t,\xi)$,  which is the circle with equation
\be
(x(\sigma)-t)^2+y(\sigma)^2 = e^{2\xi}.
\ee

From 
\be
1 =\frac{1}{y(\sigma)^2}\left\{\left(\frac{d x(\sigma)}{d\sigma}\right)^2+ \left(\frac{d y(\sigma)}{d\sigma}\right)^2\right\}
\ee
 we see that  $\sigma$ is the hyperbolic arc length. 

As a coset, the space of geodesics  inherits a metric from the group. In the $t,\xi$  parameterization the metric is
\be
ds^2= d\xi^2 - e^{-2\xi} dt^2.
\ee
With $y= e^\xi$ this becomes 
\be
ds^2= \frac{-dt^2 + dy^2}{y^2}
\ee
which is  the   AdS$_2$, or dS$_2$, metric in the Poincar\'e patch. 

The Poincar\'e patch covers only  one-half of AdS$_2$. What happened to the other half?  Although we were able obtain all the geodesics {\it as sets of points\/} by our group acting  on the reference geodesic,  the flows parameterized by $\sigma$  all went from left to right. If we want all elements in the space of {\it oriented\/} geodesics we must include  flows from right to left.  These  we get by sending  $\sigma\to  -\sigma$. This   gives us the other half, and shows that  AdS$_2$ is actually the space of {\it oriented\/} geodesics

\subsection{Isometries of the Poincar\'e disc}

The  M\"obius  map acting on $Z=X+iY$ with 
\be
Z\mapsto \frac{\lambda Z+\mu}{\bar \mu Z +\bar \lambda}, \quad |\lambda|^2-|\mu|^2=1.
\ee
preserves the  Poincar\'e disc metric 
 \be
 ds^2 =\frac{4 }{(1-X^2-Y^2)^2} (dX^2+dY^2).
 \ee
The corresponding matrix is 
\be
 g=\left(\begin{matrix}\lambda &\mu\cr \bar\mu  & \bar \lambda\end{matrix}\right)\in {\rm SU}(1,1).
\ee
The existence of a bijection from $H_+^2$ to $D^2_P$  shows that  ${\rm SU}(1,1)$ is isomorphic to ${\rm SL}(2,{\mathbb R})$.

The infinitesimal map
\be
 \left(\begin{matrix}1+\delta \lambda &\delta \mu\cr \delta \bar\mu  &1+ \delta \bar \lambda\end{matrix}\right)
\ee
can be expanded in terms of the traceless matrices \cite{kitaev}
\bea
\Lambda_0 &=&\frac 12 \left(\begin{matrix}i &0\cr 0&-i\end{matrix}\right), \nonumber\\
\Lambda_1 &=&\frac 12 \left(\begin{matrix}0 &1\cr 1&0\end{matrix}\right),\nonumber\\
\Lambda_2 &=&\frac 12 \left(\begin{matrix}0&i\cr -i&0\end{matrix}\right),
\eea
whose  algebra is
\be
[\Lambda_1,\Lambda_2]= - \Lambda_0, \quad [\Lambda_0, \Lambda_1]=\Lambda_2, \quad [\Lambda_0,\Lambda_2]= -\Lambda_1.
\ee
The   ${\rm SU}(1,1)$   quadratic Casimir in this matrix representation is 
\be
C_2 \equiv \Lambda_0^2- \Lambda_1^2-\Lambda_2^2= -\frac 34{\mathbb I}.
\ee

Now it is $\Lambda_0$ that is the compact generator.  Of the two non-compact generators $\Lambda_1$ is a boost in the $X$ direction leaving $-1$ and $1$ fixed and $\Lambda_2$ is a boost in the $Y$ direction  leaving $i$ and $-i$ fixed.
The infinitesimal M\"obius map gives 
\be
\delta Z= \delta \mu+(\delta \lambda-\delta \bar \lambda) Z- \delta \bar \mu Z^2.
\ee
Identifying  the real and imaginary parts in $\delta Z= \delta X+i\delta Y$ with those on the RHS gives us the individual $\delta X$ and $\delta Y$ and hence the vector fields $\delta X\partial_X +\delta Y \partial_Y$ taking
\bea
\Lambda_0 &\mapsto& -Y \frac{\partial}{\partial X} +X \frac{\partial}{\partial Y},\nonumber\\
\Lambda_1 &\mapsto& \{1- (X^2-Y^2)\} \frac{\partial}{\partial X} -2XY \frac{\partial}{\partial Y},\nonumber\\
\Lambda_2 &\mapsto& -2XY \frac{\partial}{\partial X} +\{1+(X^2-Y^2)\} \frac{\partial}{\partial Y}.
\eea
We confirm  that the three fields fix the points $Z=0$,  $Z=\pm1 $,  $Z=\pm i$, respectively.

If an element of ${\rm SU}(1,1)$ is written as  
\bea
g(\phi,\xi,\theta)&=& e^{\phi \Lambda_0}e^{\xi \Lambda_1} e^{-\theta \Lambda_0}, \quad \xi\ge0\nonumber\\
&=& \left(\begin{matrix} e^{i(\phi-\theta)/2}\cosh \xi/2 & e^{i(\phi+\theta)/2}\sinh \xi/2\cr e^{-i(\phi+\theta)/2} \sinh\xi/2 & e^{-i(\phi-\theta)/2}\cosh\xi/2\end{matrix}\right),
\eea
 the Poincar\'e disc is the orbit of this group action on its central point $Z=X+iY=0$. 
 The factor  $e^{-\theta \Lambda_0}$ leaves this point fixed, while 
\be
e^{\xi \Lambda_1}= \left(\begin{matrix} \cosh \xi/2&  \sinh \xi/2\cr \sinh \xi/2 &\cosh \xi/2\end{matrix}\right)
\ee
acts on $Z=0$ to give $\tanh \xi/2$. The  further action by $e^{\phi \Lambda_0}$ gives us the Poincar\'e disc coordinate
\be
g(\phi,\xi,\theta)(Z=0) = Z= e^{i\phi}\tanh( \xi/2).
\ee

\subsection{Poincar{\'e} disc geodesics}
\label{SUBSEC:poincare-geodesics}

\begin{figure}
\centering
\includegraphics[width=.2\textwidth]{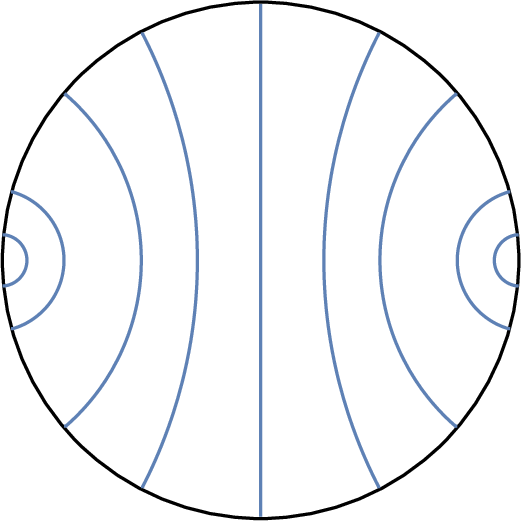}
\caption{\label{geodesics} Geodesics $(X(\sigma),Y(\sigma))$  for  various values of $\xi$.} 
\end{figure}

\begin{figure}
\centering
\includegraphics[width=.4\textwidth]{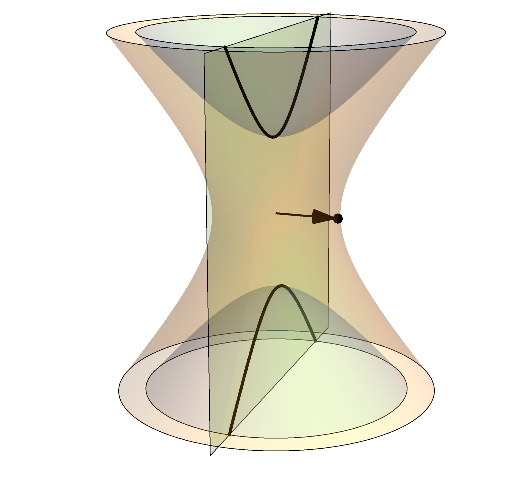}
\caption{\label{FIG:geodesicmap}
The mapping from a geodesic (the black arc)  on the upper sheet of the Minkowski-embedded hyperboloid model to a point in  de Sitter space. The arrow is the perpendicular from the center of the  plane, whose intersection with the upper hyperboloid defines  the geodesic, to the surface of the single-sheeted de-Sitter hyperboloid.  The orientation of the arrow to the left or right of the plane is determined by the direction in which the geodesic is traversed. (After figure  15 in \cite{integral-geometry})} 
\end{figure}

The geodesics on the Poincar{\'e} disc are circles that intersect the boundary at right angles. An  example is  the vertical straight line from $Z=-i$  to $Z=+i$. We can parameterize this curve by $(X(\sigma), Y(\sigma))=(0, \tanh \sigma/2)$ where $\sigma$ is the hyperbolic arc-length. It  is mapped into itself  by the isometries generated by ${\rm exp}\{ \tau \Lambda_2\}$, but is  carried into other geodesics   by ${\rm exp}\{ \xi \Lambda_1\}$. This operation provides  a family of  arc-length parametrized    geodesic curves
\be
X(\sigma)= \frac{\sinh \xi  \cosh\sigma}{1+ \cosh \xi\cosh \sigma}, \quad Y(\sigma)=  \frac{\sinh \sigma }{1+ \cosh \xi\cosh \sigma},
\ee
These curves are circles with equation
\be
(X(\sigma)- \coth \xi)^2+ Y(\sigma)^2 = \csch^2\xi. 
\ee
For positive $\xi$  the circles   intersect the boundary at points with polar angles 
\be
\tan \alpha = \left.\frac {Y(\sigma)} {X(\sigma)} \right|_{\sigma \to \infty}= \pm \frac 1 {\sinh \xi}.
\ee
We also find that  
\be
X(\sigma)^2+Y(\sigma)^2= 1-\frac{2}{1+\cosh\xi\cosh\sigma}.
\ee
As  the Euclidean distance  is related to the hyperbolic polar radius by 
\be
X^2+Y^2\equiv t^2={\rm tanh}^2 (\rho(\sigma)/2)
\ee
 we have 
\be
\cosh \rho(\sigma) =\frac{1+t^2}{1-t^2}=  \cosh{\xi}\cosh \sigma.
\ee
The polar-parametric form of the geodesic is therefore
\bea
\rho(\sigma)&=& {\rm arccosh} (\cosh{\xi}\cosh \sigma),\nonumber\\
\phi(\sigma)&=& {\rm arctan}(\csch {\xi}\, \tanh \sigma).
\eea
We can rotate this  family by an angle $\theta$, so  the  general  set of geodesics in the Poincar\'e disc constitute the  coset 
$
{\rm SL}(2, {\mathbb R})/ {\rm exp}\{ \tau \Lambda_2\}.
$
If we parameterize 
\be
g(t,\xi,\tau)= \exp\{ \theta \Lambda_0\} \exp\{ \xi \Lambda_1\}\exp\{ -\tau \Lambda_2\}
\ee
we  find the group metric  to be
\be
ds^2= 2 {\rm tr}\{ (g^{-1}dg)^2\}= d\tau^2+d\xi^2 -d\theta ^2 - 2\sinh \xi d\tau d\theta 
\ee
The metric on the coset which forgets $\tau$ is found by minimizing $ds^2$ at fixed $d\theta$, $d\xi$, so $d\tau\to \sinh \xi d\theta$, resulting in 
\be
ds^2=d\xi^2- \cosh^2\xi d\theta^2,
\label{EQ:alpha-theta-metric}
\ee
which   is a global AdS$_2$ metric. Allowing $\theta$ to range from $0$  to $2\pi$ yields  all   geodesics considered as sets {\it twice\/}--- but with opposite orientations. Figure \ref{FIG:geodesicmap} shows how this works.

\begin{figure}
\centering
\includegraphics[width=.28\textwidth]{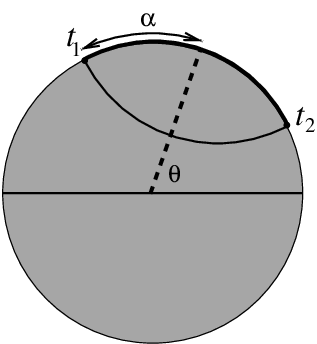}
\caption{\label{FIG:alpha-theta} The $t_1$-$t_2$ and the $\alpha$-$\theta$ parameterizations of the space of geodesics (after figure 4 in \cite{integral-geometry})} 
\end{figure}

Recall that  the parameter $\xi$ is related to (half) the angle between the endpoints of the geodesic by   $\csch\xi = \tan\alpha$.
In terms of $\alpha$  the coset  metric becomes 
\be
ds^2={\rm csc}^2 \alpha(-d\theta^2+ d\alpha^2)
\ee
which is the metric on the space of geodesics parameterized as in figure \ref{FIG:alpha-theta}.
This  is again a {\it global\/} metric on AdS$_2$ (or dS$_2$). 

\section{Spectral decompositions}
\label{SEC:eigen}
 
 \subsection{Hyperbolic Laplace eigenfunctions}
The curved-space scalar Laplacian is 
\be
\nabla^2= \frac{1}{\sqrt g}\frac{\partial}{\partial x^\mu} \sqrt g g^{\mu\nu} \frac{\partial}{\partial x^\nu}.
\ee
In Poincar\'e-disc coordinates this becomes 
\be
\nabla^2  = \frac 14 (1-X^2-Y^2)^2 \left(\frac{\partial^2}{\partial X^2}+\frac{\partial^2}{\partial Y^2}\right).
\ee
In  the upper-half-plane coordinates   
it  is 
\be
\nabla^2 =  y^2  \left(\frac{\partial^2}{\partial x^2}+\frac{\partial^2}{\partial y^2}\right),
\ee
and in  hyperbolic polar coordinates we have 
\bea
\nabla^2 &=&\frac 1{\sinh \rho} \frac{\partial }{\partial \rho} \sinh \rho  \frac{\partial }{\partial \rho}+\frac{1}{\sinh^2 \!\rho} \frac{\partial^2 }{\partial \phi^2}\nonumber\\
&=& \frac{\partial}{\partial \cosh \rho} \sinh^2 \!\rho \frac{\partial }{\partial \cosh \rho}+\frac{1}{\sinh^2 \!\rho} \frac{\partial^2  }{\partial \phi^2}.
\eea

 Formally  $-\nabla^2$  is a self-adjoint positive operator. As  Lobachevsky  space is non-compact,  it will    possess  a complete set of  eigenfunctions with a continuous spectrum. 

By analogy with the spherical harmonics  we anticipate that the  group  ${\rm PSL}(2,{\mathbb R})\simeq {\rm SO}^+(2,1)$ with $ {\rm SL}(2,{\mathbb R})\simeq {\rm SU}(1,1)\simeq {\rm Sp}(2,{\mathbb R})$ will act on the set of  functions with a given value of the Laplace eigenvalue. This is so, and each infinitely-degenerate energy level furnishes an irreducible  group representation that is one of    the  integer spin cases of    Bargmann's {\it Principal   continuous series\/} of representation of the non-compact Lie group \cite{bargmann}.

\subsubsection{The upper half-plane}
\label{SUBSUBSEC:upper}

 To find the Laplacian eigenvalues and eigenfunctions we begin with the upper-half-plane coordinate system.
Here $-\nabla^2$  is formally self-adjoint with respect to the $L^2[H_+^2]$ inner-product
\be
\brak{\phi}{\chi}= \int_{H_+^2} \phi^*(x,y) \chi(x,y) \sqrt{g}dxdy= \int_{y>0} \phi^*(x,y) \chi(x,y) \frac{dxdy}{y^2}
\ee
The Laplace operator   has $x$-translation invariance so  it is natural to seek  eigenfunctions of the form 
 $ \chi(x,y)=e^{ikx}f(y)$. 
 
 Consider first the case $k=0$. 
 The eigenvalue equation  $
-\nabla^2 \varphi=\lambda\varphi$  becomes  
 \be
 -y^2 \frac{d^2}{dy^2}\chi= \lambda \chi
 \ee
 and has a regular singular point at $y=0$. We immediately see that $y^s$ is an  eigenfunction with $\lambda = -s(s-1)$, but  what values of $s$ are suitable?    
 
 If we take $s= \frac 12 +i\kappa$, $\kappa \in{\mathbb R}$, a    complete set of  $k=0$ eigenfunctions is 
\be
\chi_{\kappa}(y)= y^{1/2+i\kappa}
\ee
with 
$
\lambda= (\kappa^2+ {\textstyle \frac 14}).
$
 The  orthogonality condition is 
\bea
\int_0^\infty \chi^*_{\kappa}(y) \chi_{\kappa'}(y)\frac{dy}{y^2}&=&\int_0^\infty y^{1/2-i\kappa}y^{1/2+i\kappa'} \frac{dy}{y^2}\nonumber\\
&=& \int_0^\infty y^{i(\kappa-\kappa')} \frac {dy}{y}\nonumber\\
&= & \int_{-\infty}^{\infty}e^{i(\kappa-\kappa')\xi}d\xi\nonumber\\
&=& 2\pi \delta(\kappa-\kappa'),
\eea
and the 
corresponding completeness relation is 
\bea
\int_{-\infty}^\infty \frac{d\kappa}{2\pi}  \chi_{\kappa}(y)\chi^*_{\kappa}(y')&=&\sqrt{yy'}  \delta(\ln y-\ln y'),\nonumber\\
&=& yy' \delta(y-y').
\label{EQ:half-plane-delta}
\eea
A substitution $y=e^\xi$, $y'= e^{\xi'}$ shows that
\be
\int_0^\infty  \sqrt{yy'}  \delta(\ln y-\ln y') f(y')\frac{dy'}{y'^2}= f(y),
\ee
so the RHS of (\ref{EQ:half-plane-delta})  is indeed the appropriate hyperbolic Dirac delta-function.

The case with $k\ne 0$ is more complicated. 
The eigenvalue equation becomes 
\be
 - \frac{d^2\chi}{dy^2}+ \left(k^2 - \frac{ \lambda}{y^2}\right)  \chi=0, 
 \ee
 which, with  $\lambda = -(\nu^2-{\textstyle \frac 14})$, would have  solutions $\sqrt y J_{\pm \nu}(ky)$.  However $-\nabla^2$ is a positive operator so  the eigenvalue $\lambda$ should   be positive. With $\lambda>0$ the  solutions will oscillate for small $y$ but   either decay or grow rapidly once  $yk> \sqrt{\lambda}$. The normalizable eigenfunctions will be those solutions that decay.
 
 There is a powerful strategy   that exploits  the  ${\rm SL}(2,{\mathbb R})$ symmetry to find  an   expression for the general-$k$ eigenfunctions as an integral. We illustrate it in detail here because we will need  to refer to it later in several places.  
 
 We  
use $y^s= ({\rm Im}(z))^s$ being an eigenfunction of $-\nabla^2$ with eigenvalue $-s(s-1)$ together with the fact that the  Laplacian commutes with the map $z\to -1/z$ to see that 
\be
\left({\rm Im}\left(-\frac 1{z}\right)\right)^s= \frac{y^s}{(x^2+y^2)^s}
\ee
remains  an eigenfunction of $-\nabla^2$ with eigenvalue $-s(1-s)$. This eigenvalue equals $\kappa^2+{\textstyle \frac1 4} $ when $s= {\textstyle \frac 12 }+i\kappa$.  Then,  because  $-\nabla^2$ commutes with  $z\to z+u$, 
 we can  form linear combinations and construct eigenfunctions  
\bea
\chi_{k, \kappa}(x,y) &\propto& \int_{-\infty}^{\infty} \left( {\rm Im}\left(\frac{1}{u-z}\right)^{\!\!s}\right) e^{iku} du, \quad s= i\kappa + {\textstyle \frac 12},\nonumber\\
&=&\int_{-\infty}^{\infty}  \frac{y^s}{((x-u)^2+y^2)^s} e^{iku} du,\nonumber\\
&=&e^{ikx} y^s \int_{-\infty}^{\infty}  \frac{1}{(u^2+y^2)^s} e^{iku} du, \quad y>0,
\label{EQ:powerful-trick}
\eea
with the property $\chi_{k, \kappa}(x+a,y)=e^{ika} \chi_{k, \kappa}(x,y)$.
With $s= \nu+\textstyle{\frac 12} $ this expression is proportional to   $e^{ikx}\sqrt{y} K_\nu(|k|y)$ where $K_\nu(y)$ is   Basset's expression 
\bea
K_\nu(y)&=& \Gamma(\nu+{\textstyle \frac 12}) \frac{(2y)^{\nu}}{2\sqrt\pi} \int_{-\infty}^{\infty} \frac{e^{iu}}{(y^2+u^2)^{\nu+1/2}} du, \quad {\rm Re}(\nu)>-1/2, \quad |{\rm Arg}(y)|<\pi/2 \nonumber\\
&=&\Gamma(\nu+{\textstyle \frac 12}) \frac{(2y^{-1})^{\nu}}{2\sqrt\pi} \int_{-\infty}^{\infty} \frac{e^{iyu}}{(1+u^2)^{\nu+1/2}} du
\eea
 for the   MacDonald function --- {\it i.e.} the modified Bessel function.

The more familiar  definition  
\bea
K_\nu(y)&=&\frac12 \int_0^\infty \exp\left\{ -\frac y2 \left( u+\frac 1 u\right)\right\} u^{\pm \nu-1}du,\nonumber\\
&=& \frac12 (y/2)^{-\nu} \int_0^\infty \exp\left(- s- \frac{y^2}{4s}\right) s^{\nu-1}{ds} 
\eea
can be recovered from Basset's by writing 
\be
\Gamma(\nu+{\textstyle \frac 12})\frac 1 {(1+u^2)^{\nu+\textstyle{\frac 1 2}}} = \int_0^\infty s^{\nu+\textstyle{\frac 1 2}} e^{-s(1+u^2)} \frac{ds}{s}
\ee
and evaluating the resulting Gaussian integral over $u$
\bea
I&=&\int_{0}^{\infty}\left(\int_0^\infty \frac {ds}{s} s^{\nu+\textstyle{\frac 1 2}} e^{iy u} e^{-s(1+u^2)}\right)du\nonumber\\
&=&\frac{ \sqrt{\pi}}{2}\int_0^\infty {ds} s^{\nu-1}\exp\left\{-s-\frac {y^2}{4s}\right\}.
\eea
With purely imaginary index $\nu=i\kappa$, 
\be
 \left(- \frac{d^2\chi}{dy^2}+ k^2 - \frac{ \kappa^2+1/4}{y^2}\right)\sqrt{y} K_{i\kappa}(|k|y)=0,
 \ee 
so the eigenvalue is again $\lambda= ({\textstyle\frac 14 +\kappa^2})$.  The  functions 
 \bea
\chi_{k,\kappa}(x,y) &=& e^{ikx}\sqrt y K_{i\kappa}(|k| y),\quad -\infty<k<\infty, \quad \nu >0
\eea
constitute    a complete set of eigenfuctions for $-\nabla^2$.  

As  
\be
{\rm K}_{i\kappa}(y) = \frac 12 \int_{0}^{\infty}\exp\left\{ - \frac{y}{2} \left(t+\frac 1 t\right)\right\}t^{i\kappa-1} dt= \frac 12 \int_{-\infty}^{\infty} \exp\left\{ - y \cosh \xi\right\}e^{i\kappa \xi}d\xi.
\ee
  these functions   are real-valued  when  $0<y<\infty$, and obey ${\rm K}_{i\kappa}(y)={\rm K}_{-i\kappa}(y)$.  They decay exponentially for $y>\kappa$ and oscillate  infinitely many    times  between $y=\kappa$, and  $y=0$.

\begin{figure}
\centering
\includegraphics[width=.7\textwidth]{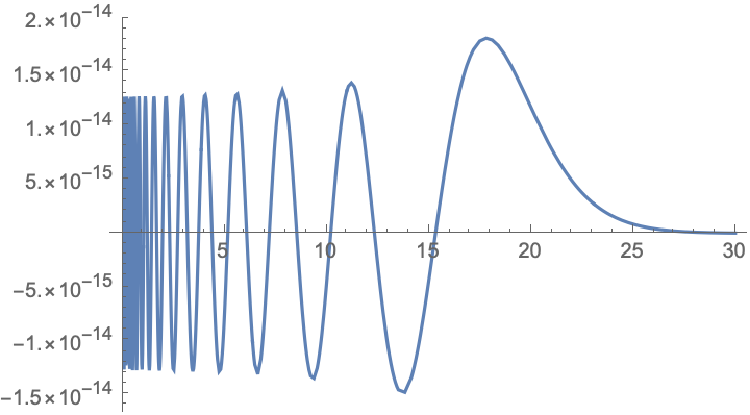}
\caption{\label{FIG:BesselK20I} A  plot of $K_{i\kappa}(y)$ for $\kappa=20$.}
\end{figure}
For large positive $x$ the leading behaviour is
\be
K_\nu(x)\sim \sqrt{\frac \pi{2x}} e^{-x}
\ee
independently  of $\nu$. 
For small positive $x$
\bea
{\rm K}_{i\kappa}(x)&\sim& \frac{i\pi}{2\sinh \pi \kappa}\left\{ \frac{(x/2)^{i\kappa}}{\Gamma(1+i\kappa)}- \frac{(x/2)^{-i\kappa}}{\Gamma(1-i\kappa)}\right\}\nonumber\\
&=&i \sqrt\frac{\pi}{\kappa \sinh \pi \kappa} \left\{ e^{i\alpha(\nu)} (x/2)^{i\kappa} - e^{-i\alpha(\kappa)} (x/2)^{-i\kappa}\right\}
\label{EQ:small-x-K}
\eea
for some real $\alpha(\kappa)$. We have used, at the last step,
\be
|\Gamma(1+i\kappa)|^2=\Gamma(1+i\kappa)\Gamma(1-i\kappa)= \frac{\pi \kappa}{\sinh \pi \kappa}.
\ee

The small-$x$ formula (\ref{EQ:small-x-K}) together with 
\be
\int_{-\infty}^\infty |x|^{i\kappa} |x'|^{-i\kappa}d\kappa = 2\pi \delta(\ln |x|-\ln |x'|)=2\pi  x\delta(x-x'), \quad x,x'>0,
\ee
is sufficient to show that these functions satisfy a completeness condition
\be
\frac{1}{\pi^2} \int_0^\infty 2\kappa\sinh \kappa\pi \,{\rm K}_{i\kappa}(x){\rm K}_{i\kappa}(x')\,d\kappa= x \delta(x-x').
\ee
The small-$x$ formula may only appear give a  completeness relation valid for small $x$,  but it reveals that    the $x$-independent normalization factor  is $(2\kappa/\pi^2)\sinh \kappa\pi $. As there are no bound states to be added to the completeness relation,   the condition must hold for   all $x$.  

The completeness relation  implies 
the orthogonality property
\be
\frac{1}{\pi^2}\int_0^\infty \frac{dx}{x} {\rm K}_{i\mu}(|k|x){\rm K}_{i\nu}(|k|x)= \frac{\delta(\mu-\nu)}{2\nu \sinh \nu \pi}.
\ee
We therefore find that the normalized eigenfunctions
\be
\chi_{k,\kappa}(x,y) =  \frac 1 \pi \sqrt{2 \kappa\sinh \pi \kappa}\, e^{ikx}\sqrt y K_{i\kappa}(|k| y)
 \ee
 satisfy both
\be
-\nabla^2\chi_{k,\kappa}(x,y)= \left(\kappa^2 +\textstyle{\frac 14}\right)\chi_{k\kappa}(x,y),
\ee
and the  the orthogonality condition
\bea
\brak{\chi_{k,\kappa}}{\chi_{k',\kappa'}}&=&\int_{-\infty}^{\infty}dx \int_0^\infty \frac{dy}{y^2} \chi^*_{k,\kappa}(x,y)  \chi_{k',\kappa'}(x,y)
\nonumber\\
&=& 2\pi  \delta(k-k')\delta(\kappa-\kappa').\
\eea

The orthogonality properties of  $K_{i\nu}$  give us the  {\it Kontorovich-Lebedev\/}  transform pair
\bea
\tilde f(\nu)\equiv K[f](\nu)&=& \int_0^\infty {\rm K}_{i\nu}(x) f(x)\,dx,\nonumber\\
f(x)&=& \frac{1}{\pi^2 x} \int_0^\infty 2\nu \sinh \nu\pi\, {\rm K}_{i\nu}(x) \tilde f(\nu)\,d\nu.
\eea

\subsubsection{Hyperbolic polar coordinates}

 Here the Laplacian  is 
\bea
\nabla^2 \varphi &=&\frac 1{\sinh \rho} \frac{\partial }{\partial \rho} \sinh \rho  \frac{\partial \varphi }{\partial \rho}+\frac{1}{\sinh^2 \!\rho} \frac{\partial^2 \varphi }{\partial \phi^2},\nonumber\\
&=& \frac{\partial}{\partial \cosh \rho} \sinh^2 \!\rho \frac{\partial\varphi }{\partial \cosh \rho}+\frac{1}{\sinh^2 \!\rho} \frac{\partial^2 \varphi }{\partial \phi^2},
\eea
and $-\nabla^2$ is self-adjoint with respect to the inner product
\be
\brak{\phi}{\chi} = \int_0^{2\pi}\!\! \int_0^\infty \phi^*(\rho,\phi) \chi(\rho,\phi)\sinh\rho \,d\rho d\phi
\ee

Separating variables as 
\be
\varphi(\rho,\phi)= \chi(\cosh \rho) e^{im\phi}, \quad m\in {\mathbb Z},
\ee
and setting $x=\cosh \rho$ makes the eigenvalue equation $-\nabla^2\varphi=(\kappa^2+ \textstyle{\frac 14})\varphi$   into a version 
\be
-\frac{d\chi}{dx}(x^2-1) \frac {d\chi}{dx} +\frac{m^2}{x^2-1} \chi = (\kappa^2+ \textstyle{\frac 14})\chi
\ee
of the associated Legendre equation, but  for the less usual range $x\ge 1$.
The $m=0$ solutions that are finite at $x=\cosh \rho=1$ are the {\it Conical functions\/} 
$
\varphi_\kappa(x)
\equiv  P_{i\kappa-\textstyle{\frac 12} }(x)
$.  They  are  the solutions to the differential equation
\be
\frac{d}{dx}(x^2-1)\frac{d}{dx} \varphi_\kappa + (\kappa^2 +\textstyle{\frac 14})\varphi_\kappa=0
\ee
in the interval $[1,\infty]$ that obey the  boundary condition $\varphi_\kappa(1)=1$.  The $\varphi_\kappa(x)$   are real-valued when $\kappa^2$ is real and positive,  $\varphi_\kappa(x)=\varphi_{-\kappa}(x)$ and the   obey the  orthogonality condition
\be
\int_1^\infty \varphi_\lambda(x)\varphi_\mu(x)\,dx= \frac{1}{\lambda \tanh (\pi \lambda)}\delta(\lambda-\mu), \quad\lambda,\mu>0,
\ee

The set $\{\varphi_\kappa: \,\, 0\le \kappa<\infty\}$ is  complete in    $L^2[1,\infty]$, and the invertible {\it Mehler-Fock transform} $\tilde f(\lambda) $ of a function $f(x)$ defined on $[1,\infty]$ is given by
\bea
\tilde f (\lambda) &\stackrel{\rm def}{=}& \lambda\, \tanh \pi  \lambda\int_1^{\infty}\varphi_\lambda(x)f(x)\,dx,\nonumber\\
f(x)&=&\int_0^\infty \phi_\lambda (x) \tilde f(\lambda)d\lambda.
\eea

\begin{figure}
\centering
\includegraphics[width=.6\textwidth]{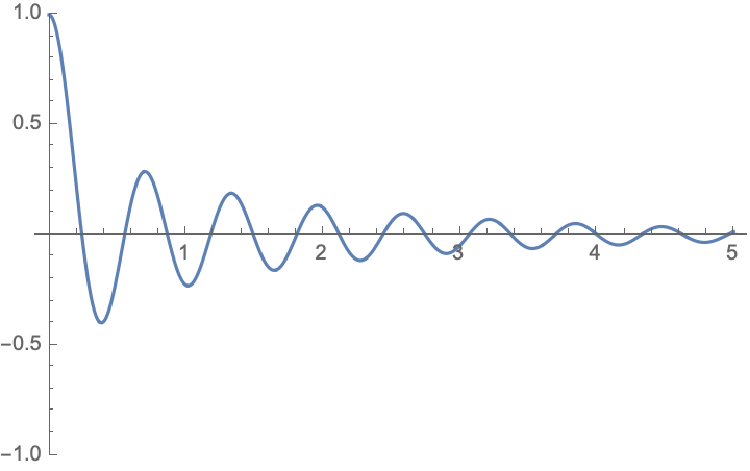}
\caption{\label{FIG:conical}  A  plot of $P_{ i\kappa-{\textstyle \frac 12}}(\cosh \rho)$ versus $\rho$ for $\kappa=10$.}
\end{figure}

The  general-$m$ eigenfunctions are
\be
\varphi_{m,\kappa}(\rho,\phi) = e^{im\phi} P^m_{i\kappa-{\textstyle \frac 12}}(\cosh \rho), \quad m\in {\mathbb Z}, \quad \kappa>0,
\ee
where $P^m_{i\kappa-{\textstyle \frac 12}}(\cosh \rho)$ is the Hobson-definition associated  Legendre function. Because, for positive integer $m$, the usual definition of the associated Legendre function is
\be
P^m_\nu(x) = (-1)^m  (1-x^2)^{m/2} \frac{d^m}{dx^m} P_\nu(x)
\ee
and the factor of $(1-x^2)^{m/2}$ means that for $m$ an odd integer these functions have   a cut which is often  taken to be along the $|x|>1$ part of real axis, which is where exactly where we need to evaluate. In particular the  Mathematica implementation makes $ P^m_{i\nu-1/2}(\cosh \rho)$ purely imaginary for $m$ odd. For us it is more convenient to relocate the  cut so the   $x>1$ functions become real. 

\begin{figure}
\centering
\includegraphics[width=.6\textwidth]{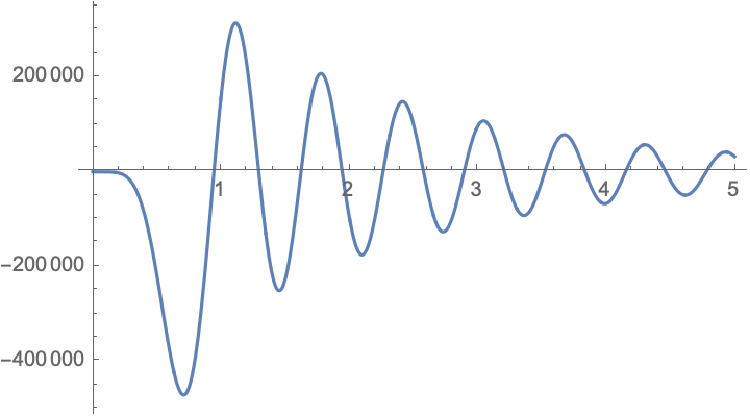}
\caption{\label{FIG:assoc_conical} A  plot of $P^6_{ i\kappa-{\textstyle \frac 12}}(\cosh \rho)$ versus $\rho$ for $\kappa=10$.}
\end{figure}

At large positive $x$, we then have 
\be
P^{m}_{i\kappa-1/2}(x)\sim \frac 1{\sqrt{2 \pi x}}\left(\frac{\Gamma(i\kappa)}{\Gamma(1/2+i\kappa -m)} (2 x)^{i\kappa} + \frac{\Gamma(-i\kappa)}{\Gamma(1/2-i\kappa  -m)}  (2x)^{-i\kappa} \right),
\ee
and the  orthogonality condition derived from  this is 
\be
\int_{0}^\infty P^m_{i\mu-1/2}(\cosh \xi) P^m_{i\nu -1/2} (\cosh \xi) \sinh \xi d \xi
= \frac{\pi}{\nu \sinh\pi \nu} \frac {\delta(\mu-\nu)}{\Gamma({\textstyle \frac 12}+i \nu-m)\Gamma({\textstyle \frac 12}-i \nu-m)}.
\ee
Taking into account the $e^{i m \theta}$ factor we have 
\be
\brak{\varphi_{n,\kappa}}{\varphi_{n',\kappa'}}= \frac{2\pi^2 \delta_{nn'} \delta(\kappa-\kappa')}
{\kappa \sinh \pi \kappa\, \Gamma(1/2-n+i\kappa)\Gamma(1/2-n-i\kappa)}.
\ee
The identity  
\be
\Gamma(1/2 +i\kappa)\Gamma(1/2-i\kappa)= \pi \,\sech (\pi \kappa).
\ee
shows that  this general-$m$ normalization is consistent with the previous $m=0$ case.

For integer $m$ but all $\alpha$ we have
\be
P_{\alpha}^m(\cosh \rho ) = \frac{\Gamma(\alpha+m+1)}{2\pi \Gamma(\alpha+1)} \int_0^{2\pi} (\sinh \rho\cos\theta +\cosh \rho)^{\alpha}e^{im\theta}d\theta.
\label{EQ:powerfultrick2}
\ee
From the formula 
\be
y= \frac 1{\cosh \rho-\cos\phi \sinh \rho}
\ee
relating the $y$ upper-half-plane coordinate to the Poincar\'e disc polar coordinates
we recognize (\ref{EQ:powerfultrick2})  as a weighted sum of the $k=0$ half-plane solutions $y^{-\alpha}$  over the orbit around the centre of the Poincar\'e disc. It therefore  arises by the same strategy that gave the half-plane eigenfunctions.    

\subsection{Liouville Potential}
\label{SUBSEC:liouville}

 As we have seen in section \ref{SUBSUBSEC:upper},  the Liouville-potential      Schr{\"o}dinger equation
\be
\left(-\frac{d^2}{d\xi^2}+m^2 e^{2\xi}\right) \psi=E\psi(\xi)
\ee
possesses a purely continuous spectrum with $\delta$-normalized eigenfunctions 
\be
\psi_\nu(\xi)=\left( \frac{2\nu \sinh  \nu \pi}{\pi^2}\right)^{1/2} K_{i\nu}(|m| e^{\xi}), \quad E= \nu^2, 
\ee   
\be
\int_{-\infty}^{\infty} \psi_\mu(\xi)\psi_\nu(\xi)\, d\xi = \delta(\nu-\mu).
\ee

Changing the sign  $m^2\to -k^2$ leads to    the    Schr{\"o}dinger operator
\be
H=-\frac{d^2}{d\xi^2}-k^2 e^{2\xi},
\ee
which appears in \cite{joeP}, and 
in which the potential  is unbounded below. 
Comparing with Bessel's equation 
\be
\left(x^2\frac{d^2}{dx^2}+ x \frac{d}{dx }+ x^2 -\nu^2\right) \psi=0
\ee
and writing $x=e^\xi$, $E=-\nu^2$ shows us that
the differential equation $H\psi=E\psi$ has solutions 
\be
\psi_\nu(\xi)= J_{\pm \nu}(|k| e^{\xi}),\quad E=-\nu^2.
\ee
The problem is to   select  from these  solutions a complete orthonormal set.

Lagrange's identity for the Liouville  problem
\be
\int_a^b  (\psi^* H\chi- (H\psi)^* \chi)d\xi= [ \psi^* \partial_\xi \chi - (\partial_\xi \psi)^* \chi]_a^b, \quad a,b\in {\mathbb R},
\ee
 translates to the  Bessel integral
\be
(\beta^2-\alpha^2)\int_a^b \frac{dx}{x} J_\alpha(|\mu|x) J_\beta(|k|x) = \left[x\left(J_\alpha (|k|x) \partial_x J_\beta(|k|x) -  (\partial_x J_\alpha (|\mu|x)) J_\beta(|\mu| x)\right)\right]_a^b, \quad a,b>0.
\ee
Near zero, and for any value of $\alpha$ except negative integers,  
\be
J_\alpha(x) \sim \frac 1{\Gamma(\alpha+1)} x^{\alpha}= \frac 1{\Gamma(\alpha+1)}e^{\alpha \xi}. 
\ee
For large $|x|$, $|{\rm Arg}(x)|<\pi$ we have
\be
J_{\alpha}(x) \sim \sqrt{\frac{2}{\pi x}} \cos\left(x-\alpha\pi/2-\pi/4\right).
\ee
We see that  for $\alpha, \beta $ real and positive, the contribution at zero vanishes while  that at infinity has a finite limit \cite{joeP}. The result is 
\be
\int_0^\infty \frac{dx}{x}J_\alpha (k|x) J_{\beta}(|k| x) = \frac 2\pi \frac{\sin\left(\frac \pi 2 (\beta-\alpha)\right)}{\beta^2-\alpha^2}.
\label{EQ:Jn-orthogonality}
\ee
The RHS of (\ref{EQ:Jn-orthogonality})  is zero if $\alpha$ and $\beta$ differ by twice an integer.   The limit $\alpha\to \beta$ is also finite. So,  quite remarkably,  we find that for any real number $\theta$ and with  $\nu_n = 2(n+\theta/\pi)$,  $n\in {\mathbb Z}$ we have found   a square-integrable  set obeying 
\be
 \int_0^\infty \frac{dx}{x}J_{\nu_n} (|k|x) J_{\nu_{n'}}(|k| x) =  \frac{\delta_{nn'}}{2\nu_{n'}},
 \ee
 or equivalently
 \be
 \int_{-\infty}^\infty {d\xi }\,J_{\nu_n} (|k|e^\xi) J_{\nu_{n'}}(|k| e^\xi) =  \frac{\delta_{nn'}}{2\nu_{n'}}.
 \ee
 We therefore have  a collection  of  families, parametrized by the angle  $\theta$ ${\rm mod}\, \pi$, each containing infinitely many  normalizable and mutually orthogonal  states with energy $-\nu_n^2$. How should we select among the families?  And what about positive energy states?

 This is a classic problem of  seeking   a domain on  which an  operator is self-adjoint.
 We  must first define  a minimal domain on which the $H$ operator is symmetric
 \be
 \brak{\psi}{H\phi}= \brak{H\psi}{\phi},
 \ee
 and then seek to extend it.
 A minimal domain is one in which  we require that both the functions and their derivatives  vanish at the boundaries.  Examination  of the solutions  of $(H\pm i)\psi=0$ on this domain reveals  that both signs of $i$ have a single normalizable solution, so the deficiency indices are $n_+=n_-=1$.  The  Weyl-Von Neumann theorem now tells us that there is a one-parameter family of self-adjoint extensions of $H$. To find these we begin by observing  that the  point $x=\infty$ is a singular endpoint of  limit-circle  type. Near such a point there are two linearly independent solutions  whose behaviour is independent of the eigenvalue.  
 A choice of self-adjoint extension is made by requiring that {\it all\/}  solutions be a fixed  linear combination of these two solutions.   In the present case we do this by  imposing the condition that  both positive and negative  energy wavefunctions  behave as
 \be
 \psi \sim \sqrt{\frac{2}{\pi x}} \cos\left(x-\theta-\pi/4\right), \quad x\to +\infty,
 \ee
 where $\theta$ is the extension parameter.
  This condition is equivalent to imposing a  Dirichlet boundary condition  at some distant $\xi=L$ that   requires  the argument of the cosine there to be an odd multiple of $\pi/2$, so $\theta$ is related to $L$ by
\be
\theta =e^{L}- \frac{3\pi}{4}, \quad {\rm mod} \,\pi.
\ee
In most Schr\"odinger-potential problems   a far-away wall  has negligible influence on the spectrum.  The quantum particle  has to actually reach the wall and return before we can see  its effect. When the round trip  takes  a  long time the   ``energy-time uncertainty principle''   means  that  we must  examine the energy spectrum on a very fine scale to detect  the wall's influence.  In the present case  the singular potential allows the  particle to  travel to the  wall, bounce off it and return all  in a \underline{finite} time.  The wall is therefore able to control the spectrum.

It is convenient to define \cite{dunster90}
\bea
 F_\nu(x) &=&\frac 12 \sec\!\left( \frac {\nu\pi}{2}\right) (J_\nu(x) +J_{-\nu}(x)),\nonumber\\
 G_\nu(x) &=&\frac 12 \cosec\!\!\left( \frac {\nu\pi}{2}\right) (J_\nu(x) -J_{-\nu}(x)). 
\eea
For the scattering states with positive energy $E=\kappa^2$ we take $\nu=i\kappa$ and find that as $\xi\to -\infty$ these functions become plane waves
\bea
 F_{i\kappa}(e^\xi) &\sim& \sqrt{\frac{2\tanh \pi \kappa/2}{\pi \kappa}} \cos(\kappa\xi-\delta_\kappa),\nonumber\\
 G_{i\kappa}(e^\xi) &\sim& \sqrt{\frac{2\coth \pi \kappa/2}{\pi \kappa}} \sin(\kappa\xi-\delta_\kappa),
\eea
where $\delta_\kappa$ is an unimportant (for us) phase shift.

As $\xi\to +\infty$ 
\bea
 F_{i\kappa}(e^\xi) &\sim& \sqrt{\frac{2}{\pi }} e^{-\xi/2}\cos(e^\xi-\pi/4),\nonumber\\
 G_{i\kappa}(e^\xi) &\sim& \sqrt{\frac{2}{\pi }} e^{-\xi/2}\sin(e^\xi-\pi/4),
\eea
where the right-hand sides are   independent of $\kappa$.
The   $\theta$-dependent large-$\xi$ boundary condition is  satisfied by taking
\be
 \Xi_\kappa(\xi) = {\mathcal N}_\kappa \left(F_{i\kappa}(e^\xi)\cos\theta + G_{i\kappa} (e^\xi)\sin \theta\right),
\ee
where the  normalization factor
\be
{\mathcal N}_\kappa^{-2}=\frac 1 {\kappa}\left(\tanh(\pi \kappa/2) \cos^2 \theta + \coth (\pi \kappa/2) \sin^2\theta\right)
\label{EQ:liouville-norm}
\ee
is derived  in \cite{andrianov} eqn.\ 23\footnote{Beware of a typo: while their derivation from  their equations 12 and 13 is correct, in writing down eq 23 the authors of \cite{andrianov} have  accidentally squared the tangent, cotangent and $1/\kappa$.}.

 The $E= -\nu^2$ negative-energy bound states  are given by taking real $\nu$
\bea
\Phi_\nu(\xi) &=&F_{\nu}(e^\xi)\cos\theta + G_{\nu} (e^\xi)\sin \theta\nonumber\\
&\propto&J_\nu(e^\xi) \sin\left(\frac{\pi \nu}{2}+\theta\right) + J_{-\nu}(e^\xi) \sin\left(\frac{\pi \nu}{2}-\theta\right),
\eea
with
\be
\nu\to \nu_n= 2\left(n+\frac \theta\pi  \right)
\ee
so that $\Phi(\xi)$ obeys the  large-$\xi$ boundary condition $ \psi \sim \cos\left(x-\theta-\pi/4\right)$.
This choice makes also makes  the coefficient of $J_{-\nu}(e^{\xi})$ zero, so  ensuring    that the wavefunction decays exponentially as $\xi\to -\infty$.  
The normalized bound states are then
\be
\Phi_n(\xi) = \sqrt{2(n+\theta/\pi)} J_ {2(n+\theta/\pi)}(e^\xi).
\ee
These, together with the scattering states, compose a complete  orthonormal set
\bea
\int_{-\infty}^\infty  \Phi^*_n(\xi) \Phi_{n'}(\xi)\,d\xi &=& \delta_{nn'},\nonumber\\
\int_{-\infty}^\infty  \Phi^*_n(\xi) \Xi_{\kappa}(\xi)\,d\xi &=& 0\nonumber,\\
\int_{-\infty}^\infty    \Xi^*_{\kappa}(\xi) \Xi_{\kappa'}(\xi)\,d\xi &=& \delta(\kappa-\kappa').
 \eea
 
In \cite{joeP}  a specific choice of $\theta=3\pi /4 $ is made. In this case 
\be
\Xi_\kappa(x)\to \sqrt{\frac 2 {\pi x}} \cos x, \quad x\to \infty
\ee
and the normalization factor (\ref{EQ:liouville-norm}) becomes 
\be
{\mathcal N}^{-2}_\kappa= \frac{\coth \pi \kappa}{\kappa}.
\ee

The authors of \cite{joeP}  prefer to define a function $Z_\nu(x)$ by writing
\bea
 &&F_\nu(x)\cos\theta + G_\nu(x)\sin \theta\nonumber\\
 &=&  \frac{\cosec (\nu\pi/2)-\sec(\nu\pi/2)}{2\sqrt{2}} \left(J_{\nu}(x)+\frac{\tan(\nu\pi/2)+1}{\tan(\nu\pi/2)-1}J_{-\nu}(x)\right)\nonumber\\
 &=& \frac {\cosec (\nu\pi/2)-\sec(\nu\pi/2)}{2\sqrt{2}} Z_\nu(x).
 \eea
 For purely imaginary $\nu= i\kappa$ 
 \bea 
 &&F_{i\kappa}(x)\cos\theta  +G_{i\kappa}(x)\sin\theta\nonumber\\
 &=& - \frac{\sech(\kappa \pi/2)+ \csch(\kappa \pi /2)}{2 \sqrt{2}} Z_{i\kappa}(x)\nonumber\\
 &=& -\frac{i}{\sqrt{2}} \frac{\sin(\pi({\textstyle \frac 1 4 - \frac{i\kappa}{2}}))}{\sinh \kappa \pi} Z_{i\kappa}(x),
 \eea
 and we
may use (\ref{EQ:liouville-norm})  to see that 
 \be
 \int_0^\infty Z^*_{i\kappa}(|\mu|x)Z_{i\kappa'}(|\mu|x)\frac{dx}{x} = \frac{2 \sinh(\pi \kappa)}{\kappa}\delta(\kappa-\kappa'),
 \ee
 in agreement with \cite{joeP}.
 
 \section{Radon Transforms}
 \label{SEC:radon}

A Radon or X-ray  transformation  is map from functions $f(z)$ on some Riemann manifold  $M$ to  the space $\Gamma(m)$  of geodesic curves on $M$. The map is performed    by integrating the function over the   curve
  \be
  f(z) \mapsto [{\mathcal R}f](\gamma) = \int_\gamma f(z(\sigma))d\sigma,
  \ee
  where  $\gamma(\sigma)$ is a geodesic parametrized by its  arc length. 
  $\Gamma(M)$ is an example  \cite{integral-geometry} of a 
   ``kinematic space,'' which  has a number of applications to AdS/CFT duality \cite{hyperbolic_radon_AdS,Ma1,Ma2}.  
 
  As we have seen  in section \ref{SEC:geometry} the space of geodesics on Bolyai-Lobachevski  space  can be regarded as a coset of ${\rm SL}(2,{\mathbb R})$ by a hyperbolic subgroup. As such, it inherits a metric from the group, and from the  group-Casimir  a Laplacian. This Laplacian is actually a  wave-operator $\nabla^2_{dS}$ on dS$_2$.     The key fact is that ${\mathcal R}$ intertwines the Laplacian $\nabla^2_H$ on the Euclidean signature hyperbolic space with  that  on the Lorentz signature (de-Sitter) kinematic space. In other words 
  \be
 - \nabla^2_{dS}({\mathcal R}f)= {\mathcal R} (-\nabla^2_H f).
  \ee
  An eigenfunction of  $\nabla^2_H$ is therefore transformed into an eigenfunction of $\nabla^2_{dS}$ with the same eigenvalue. The intertwining property  comes about because both the wave operator $\nabla^2_{dS}$ and the Laplacian $\nabla^2_{H}$ are the same differential operator ---  the ${\rm SL}(2,{\mathbb R})$ group Casimir --- but  acting on functions on the group that are constant along the fibres of the different  equivalence classes  that  constitute the   cosets. The Radon transform takes sums of eigenfunctions that are constant on   $H^2_+= {\rm SL}(2,{\mathbb R})/K$  equivalence classes and by averaging turns them into functions that are constant on    dS$_2$ equivalence classes. This is a special case  of the strategy  that converted $y^s$ into the general eigenfunction in section \ref{SUBSUBSEC:upper}.

 \subsection{Upper half-plane}
 \label{SUBSEC:radon-upper}
  
 Consider the case of the upper-half-plane. Parametrize an ${\rm SL}(2,{\mathbb R})$ group  element by its NHA factorization (see appendix \ref{SUBSUBSEC:NHA})
  \bea
  g&=& n^T(t) h(\xi) a(\phi)\nonumber\\
  &=&\exp\{t L_{-1}\} \exp\{ \xi  L_0\} \exp\{\phi (L_{-1}-L_1)/2)\}\nonumber\\
  &=&
 \left(\begin{matrix} e^{\xi/2} \cosh(\phi/2)+t e^{-\xi/2} \sinh(\phi/2)  & e^{\xi/2} \sinh(\phi/2)+t e^{-\xi/2} \cosh (\phi/2)  \cr
 e^{-\xi/2} \sinh (\phi/2) & e^{-\xi/2} \cosh(\phi/2)\end{matrix}\right),
  \eea
  and recall that this group element transforms our reference geodesic into another
     geodesic $\gamma_{\xi, t}$  equipped with arc-length parametrization   
   \bea
   x(\sigma)&=& e^\xi\, \tanh \sigma+t,\nonumber\\
   y(\sigma)&=&e^\xi \,\sech \sigma.\
   \eea
  The corresponding ${\rm SL}(2,{\mathbb R})/A$ coset metric is 
  \be
  ds^2= d\xi^2-e^{-2\xi}dt^2.
  \ee
  If we  write $e^\xi=\eta$ this becomes 
  \be
ds^2= \frac{-dt^2 + d\eta^2}{y^2},
\ee
 which is the AdS$_2$ metric in the Poincar\'e patch coordinates. 
 
 The Wave operator eigenvalue equation   $-\nabla^2_{dS}\psi= (\nu^2+1/4)\psi$ with  $\psi(\xi,t)=e^{ikt} f(\xi)$ is 
 \be
 \left(- \frac{d^2}{d\xi^2} + \frac{d}{d\xi}-k^2 e^{2\xi}\right) f(\xi)= (\nu^2+1/4) f(\xi).
 \ee
 It is convenient to write $f(\xi)= e^{\xi/2}\chi(\xi)$ so that the eigenvalue problem becomes 
 \be
 \left(- \frac{d^2}{d\xi^2} -k^2 e^{2\xi}\right) \chi(\xi)= \nu^2 \chi(\xi),
 \ee
 which   is  the unbounded-below case of the Liouville potential  from section \ref{SUBSEC:liouville}. There  we saw that the  Schr\"odinger operator has deficiency indices (1,1) and a hence family of self-adjoint  extensions parameterized by an angle $\theta$. The  scattering state eigenfunctions  $f(\xi)$ are linear combinations of $e^{\xi/2} J_{\pm i\nu}(|k|e^\xi)= \sqrt{\eta} J_{\pm i\nu}(|k|\eta)$ and there are also bound state eigenfunctions with $E_n= -\nu_n^2$, $\nu_n = 2(n+ \theta/\pi)$, which for $\theta=3\pi/4$ are $\nu_n=3/2, 7/2,11/2,\ldots$.

  Now consider the Radon transforms of the upper half-plane 
  Laplace eigenfunctions 
  \be
  \chi_{k, \nu}(x,y) = N_\nu  e^{ikx} \sqrt{y} K_{i\nu}(|k|y), \quad N_\nu= \frac 1\pi \sqrt{2 \nu \sinh\pi \nu}.
  \ee
  We integrate over  the path $\gamma(\sigma)=(x(\sigma), y(\sigma))$
     to find \cite{jevicki}
  \bea
   {\mathcal R}[\chi_{k, \nu}](t, \eta) &=&N_\nu   e^{ikt}\int_{-\infty}^{\infty} e^{ik x(\sigma)} \sqrt{y(\sigma)}K_{i\nu}(|k|y(\sigma)) d\sigma\nonumber\\
   &=&N_\nu   e^{ikt}\eta \int_{-\eta}^{\eta} e^{ik x } (\eta^2-x^2)^{1/4}K_{i\nu}(|k|\sqrt{\eta^2-x^2}) \frac{dx}{(\eta^2-x^2)} \nonumber\\
   &=& N_\nu   2 \eta e^{ikt}\int_{0}^{\eta} \cos (k x ) K_{i\nu}(|k|\sqrt{\eta^2-x^2}) \frac{dx}{(\eta^2-x^2)^{3/4}}. 
  \eea 
  We know from the intertwining property that the resulting function must be one of the eigenvalue $\nu^2+1/4$ eigenfunctions 
  of the wave operator  on the kinematic space. 
 We do  not, however, know which self-adjoint extension should apply. 
 To determine the extension parameter $\theta$   we  need to find   the large $\eta$ behaviour of the integral.  Fortunately 
 this is not difficult. For   $k \eta\gg 1 $ the integration path  will extend far beyond the exponentially cut-off range of the Bessel function so we can treat the semicircle  as two vertical lines at $x=\pm \eta$.  We therefore have the asymptotic behaviour
  \bea
   {\mathcal R}[\chi_{k, \nu}](t, \eta)&\sim&   N_\nu   2  e^{ikt} \cos (k \eta)\int_0^\infty \sqrt{y} K_{i\nu}(|k|y)\frac{dy}{y}\nonumber\\ 
   & =&  N_\nu  2^{-1/2}  e^{ikt} |k|^{-1/2} \cos (k \eta) \Gamma\left({\textstyle \frac 14 +\frac{i\nu}{2}}\right)\Gamma\left({\textstyle \frac 14 -\frac{i\nu}{2}}\right).
   \eea
   The last step comes from
   \bea
  I&=&  \int_0^\infty \sqrt{y} K_{i\nu}(|k|y)\frac{dy}{y}\nonumber\\
  &=& \frac 12\int_0^\infty  y^{1/2-1}
 \left( \int_{-\infty}^{\infty} \exp\left\{ - y|k| \cosh \xi\right\}e^{i\kappa \xi}d\xi\right)dy.\nonumber\\
 &=& \frac 12 \Gamma(1/2) |k|^{-1/2} \int_{-\infty}^{\infty} {\rm sech}^{1/2} e^{i\kappa \xi}d\xi\nonumber\\
 &=&  2^{-3/2}|k|^{-1/2}\Gamma\left({\textstyle \frac 14 +\frac{i\nu}{2}}\right)\Gamma\left({\textstyle \frac 14 -\frac{i\nu}{2}}\right).
\eea
  When  comparing the asymptotic $\cos (k \eta)$  factor with our earlier discussion Liouville problem  we  see  it agrees with the large-$\eta$ asymptotic behaviour only when  $\theta = 3\pi/4$, as found in  
 \cite{jevicki}. We have in this way  evaluated the Bessel integral without having to search through Watson \cite{watson}. 
  
 After including the normalization factors $N_\nu$ and ${\mathcal N}_\nu$ for the Modified and unmodified Bessel functions, we find that the Radon transform of the $\delta$-normalized upper-half-plane mode functions gives $\delta$-normalized dS$_2$ functions multiplied by a factor 
 \be
 \Lambda(\nu)\equiv \frac 1{\sqrt{2 \pi}} \Gamma\left({\textstyle \frac 14 +\frac{i\nu}{2}}\right)\Gamma\left({\textstyle \frac 14 -\frac{i\nu}{2}}\right)\sqrt{\cosh \pi \nu}
 \ee
 which becomes $ \sqrt{2\pi/ |\nu|}$
 as $\nu$ becomes large. This  large-$\nu$  behaviour is the same as that of the  singular values of the Radon transform in ${\mathbb R}^2$ (the singular values being the positive square-root of the eigenvalues of ${\mathcal R}^*{\mathcal R}$). This is comforting as the hyperbolic Radon  transform should reduce the flat space case once  the wavelength of the modes becomes smaller than the scale factor of the curved space.  
 
 The suppression  by the  transform of the  high frequency modes means that ${\mathcal R}^{-1}$ is an unbounded operator and  its     inversion   an ill-posed problem. The source of the ill-posedness is the same as that of trying to extract  the initial data of a heat-diffusion problem from its later values: both diffusion and the Radon transform involve taking averages, and averaging loses information.

  \subsection{Poincar\'e disc}
  \label{SUBSEC:radon-poincare}
  
  The kinematic space corresponding to the Poincare disc is the same as that of the upper half-plane as both spaces are  a quotient of  ${\rm SL}(2,{\mathbb R})\simeq {\rm SU}(1,1)$ by a hyperbolic subgroup, but with a different parametrization.
   
  The hyperbolic subgroup  element   $\exp\{ -\tau \Lambda_2\}$  fixes the points $\pm i$ on the  boundary of the Poincar{\'e} disc, and if  we parameterize (see appendix \ref{SUBSEC:AdSviaSU11})
\be
g= \exp\{ \theta \Lambda_0\} \exp\{ \xi \Lambda_1\}\exp\{ -\tau \Lambda_2\}\in {\rm SU}(1,1)
\ee
we  find the group metric to be
\be
ds^2= d\tau^2+d\xi^2 -d\theta ^2 - 2\sinh \xi d\tau dt. 
\ee
The metric on the coset which forgets $\tau$ is 
\be
ds^2=d\xi^2- \cosh^2\xi d\theta^2,
\ee
which   is a global AdS$_2$ metric. If we define an angle $\alpha$ by $\csch \xi= \tan \alpha$ the this metric becomes 
\be
ds^2={\rm csc}^2 \alpha(-d\theta^2+ d\alpha^2)
\ee
which is the metric on the space of geodesics on the Poincar\'e disc when they are parameterized as in figure \ref{FIG:alpha-theta}.

From either the left or right invariant fields we find that the group Casimir acting on functions 
 $\psi= e^{i\lambda \theta} \psi(\xi) e^{i\mu\tau}$ is
\be
-\frac 1{\cosh  \xi}\frac{\partial}{\partial  \xi} \cosh \xi  \frac{\partial}{\partial  \xi}-\frac 1{\cosh^2\xi} (\lambda^2-\mu^2 +2\lambda\mu \sinh \xi) .
\ee
 Functions on the coset  do not depend on $\tau$ and so are of the form $\psi(\theta,\xi)= e^{ik \theta} \psi(\xi) $. The eigenproblem for them   is therefore  
\be
-\left(\frac{\partial}{\partial  \sinh \xi} (1+ \sinh^2 \xi) \frac{\partial}{\partial\sinh   \xi}+\frac {k^2} {1+\sinh ^2\xi}\right) \psi(\xi)= (\nu^2+1/4)\psi(\xi). 
\ee
This  is a form of Legendre's equation and has solutions $\psi(\xi)=P^k_{i\nu-1/2}(\pm i \sinh \xi)$ or $Q^k_{i\nu-1/2}(\pm i \sinh \xi)$ \cite{dunster2013}. Legendre functions with a purely imaginary argument   are not commonly seen. Perhaps, by   analogy with modified Bessel functions,   we might call them modified conical functions?

 The standard  definition of the associated Legendre functions (\cite{NIST} \S 14.5) 
  has 
\bea
P^{k}_{i\nu -1/2}(0) &=& \frac{2^k \sqrt{\pi}}{\Gamma\left(\frac 3 4 +\frac {i\nu}2 - \frac{k}{2} \right)\Gamma\left( \frac 34 - \frac {i\nu}2-\frac k2\right)},
\nonumber\\
\left.\frac{d}{dz}P^{k}_{i\nu -1/2}(z)\right|_{z=0} &=&- \frac{2^{k+1} \sqrt{\pi}}{\Gamma\left(\frac 1 4 +\frac {i\nu}2 - \frac{k}{2} \right)\Gamma\left( \frac{1}{4} - \frac {i\nu}{2}-\frac{k}{2}\right)},
\eea
which are both real when $k$ is real. Consequently  $P^k_{i\nu - 1/2}(z)$ is real when $z$ is real and the power series converges.  When $z=ix$ is purely imaginary, however, all the odd powers in the expansion are imaginary, so
\bea
 {\rm Re}[P^k_{i\nu-\frac 12}(ix)] &\quad& \hbox{is symmetric in $x$},\nonumber\\
{\rm Im}[P^k_{i\nu-\frac 12}(ix)]&\quad& \hbox{is antisymmetric in $x$}.
\eea
We will therefore define
\bea
 E^k_\nu(\xi) \equiv  {\rm Re}[P^k_{i\nu-\frac 12}(i\,\sinh \xi)],\nonumber\\
 O^k_\nu(\xi) \equiv  {\rm Im}[P^k_{i\nu-\frac 12}(i\,\sinh \xi)].
\eea

With these facts in mind   we can compute the Radon transforms.
Recall  that  the arc-length parametrized geodesic $\gamma_{\xi,\theta}$ is 
\bea
\rho(\sigma)&=& {\rm arccosh}(\cosh\sigma \cosh \xi)\nonumber\\
\phi(\sigma)&=&{\rm arctan}(\csch\xi\, \tanh\sigma)+\theta,
\eea
 so  the  Radon transform of the Poincare-disc eigenfunction $\varphi_{k,\nu}(\rho,\phi)=e^{ik\phi} P^k_{i\nu-1/2}(\cosh \rho)$  is 
\be
{\mathcal R}[\varphi_{k,\nu}](\xi,\theta)=\int_{-\infty}^\infty P^k_{i\nu-1/2}(\cosh \xi\cosh \sigma) e^{ik\phi(\sigma)}d\sigma.
\label{EQ:poincare-radon}
\ee
This is not a standard integral, but we  see  that it is symmetric or antisymmetric in $\xi$ depending on whether $k$ is an odd or even integer. It must also be an eigenfunction of the wave operator with the same eigenvalue $\nu^2+\frac 14$ and order $k$ as  the $P^k_{i\nu-1/2}(\cosh \rho)$. Therefore the  result of the integration  will  be 
\be
{\mathcal R}[\varphi_{k,\nu}](\xi,\theta)
\propto 
\begin{cases} 
e^{ik\theta}\,{\rm Re}[P^k_{i\nu-1/2}(i\sinh \xi)], & \hbox{$k$ even,}\cr
e^{ik\theta}\,{\rm Im}[P^k_{i\nu-1/2}(i\sinh \xi)], & \hbox{$k$ odd}.
\end{cases}
\ee
For the even-$k$ case  the proportionality factor   can  be found directly from the $\xi=0$ integral (\cite{oberhettinger} Inverse of third equation on page 29)
\bea
\int_{-\infty}^\infty P^k_{i\nu-1/2}(\cosh \sigma) d\sigma&=&  2^{k}
\frac{
\Gamma({\textstyle\frac 14+ \frac {i \nu} 2})\Gamma({\textstyle \frac 1 4-\frac {i \nu} 2})}{\Gamma({\textstyle \frac 3 4-\frac k 2+ \frac {i \nu} 2})\Gamma({\textstyle \frac 34-\frac k 2- \frac {i\nu} 2})}\nonumber\\
&=& \frac 1{\sqrt \pi} P^k_{\i\nu-1/2}(0) \Gamma({\textstyle\frac 14+ \frac {i \nu} 2})\Gamma({\textstyle \frac 1 4-\frac {i \nu} 2}).
\eea
We can just replace the argument ``$0$'' in the Legendre function by $i \sinh \xi$ to get
\bea
[R\varphi_{k,\nu}](\xi,\theta)&=& \int_{-\infty}^\infty P^k_{i\nu-1/2}(\cosh \sigma\cosh \xi)e^{i k \phi(\sigma)} d\sigma\nonumber\\
&=&\frac {e^{ik\theta}}{\sqrt \pi} {\rm Re}[P^k_{\i\nu-1/2}(i \sinh \xi )] \Gamma({\textstyle\frac 14+ \frac {i \nu} 2})\Gamma({\textstyle \frac 1 4-\frac {i \nu} 2}), \quad \hbox{$k$ even}.
\eea
It is now natural to conjecture that  
\bea
[R\varphi_{k,\nu}](\xi, \theta)&=& \int_{-\infty}^\infty P^k_{i\nu-1/2}(\cosh \sigma\cosh \xi)e^{i k \phi(\sigma)} d\sigma\nonumber\\
&=&\frac {e^{ik\theta}}{\sqrt \pi} {\rm Im}[P^k_{i\nu-1/2}(i \sinh \xi )] \Gamma({\textstyle\frac 14+ \frac {i \nu} 2})\Gamma({\textstyle \frac 1 4-\frac {i \nu} 2}), \quad \hbox{$k$ odd},
\eea
and this    checks numerically\footnote{When making the  numerical check we remind the reader that the Mathematica implementation makes $P^k_{i\nu - 1/2}(\cosh \xi)$ pure imaginary when $k$ is odd.}.
We  will  confirm  the conjecture by making use of   the large-$\xi$ asymptotics of $P^k_{i\nu - 1/2}(i \sinh \xi)$ from  appendix \ref{SEC:asymptotics}.

 In the large-$\xi$  limit the $e^{ik\phi(\sigma)}$ factor in \ref{EQ:poincare-radon} becomes unity, and the $ P^k_{i\nu-1/2}(\cosh \xi\cosh \sigma)$ factor can be replaced by its asymptotic form
\be
P^{\mu}_{i\nu-1/2}(z)\sim \frac 1{\sqrt{2 \pi z}}\left(\frac{\Gamma(i\nu)}{\Gamma(1/2+i\nu -\mu)} (2 z)^{i\nu} + \frac{\Gamma(-i\nu)}{\Gamma(1/2-i\nu  -\mu)}  (2z)^{-i\nu} \right).
\ee
We can also replace  $2 \cosh \xi \to e^{\xi}$.
Inserting $z\to \cosh\xi \cosh\sigma$ and integrating over $\sigma$ requires  
\bea 
\int_{-\infty}^{\infty} (\cosh \sigma)^{i\nu-1/2} d\sigma&=& \sqrt{\pi}\frac{\Gamma(1/4-i\nu/2)}{\Gamma(3/4-i\nu/2)}\nonumber\\
&=& (1/ \sqrt{\pi}) \{\Gamma(\textstyle{ \frac 14-\frac {i\nu}2})\Gamma(\textstyle{ \frac 14+\frac {i\nu}2})\} \sin[\pi(1/4+i\nu/2)], \
\eea
where in the second line we  have  re-expressed the  RHS  by using a Gamma-function identity
\be 
\sqrt{\pi}\frac{\Gamma(1/4-i\nu/2)}{\Gamma(3/4-i\nu/2)}
= (1/ \sqrt{\pi}) \{\Gamma(\textstyle{ \frac 14-\frac {i\nu}2})\Gamma(\textstyle{ \frac 14+\frac {i\nu}2})\} \sin[\pi(1/4+i\nu/2)].
 \ee
The large-$\xi$ behaviour of the Radon transform is therefore found to be $\Gamma(\textstyle{ \frac 14-\frac {i\nu}2})\Gamma(\textstyle{ \frac 14+\frac {i\nu}2})/{\pi}$ times 
\be
 \frac{\Gamma(i\nu)}{\Gamma({\textstyle \frac 12}+i\nu -k)} e^{(i\nu-1/2)\xi} \sin[\pi(1/4+i\nu/2)]+  \frac{\Gamma(-i\nu)}{\Gamma({\textstyle \frac 12}-i\nu -k)} e^{(-i\nu-1/2)\xi} \sin[\pi(1/4-i\nu/2)].
 \label{EQ:PoincareRadonAsymptotic}
 \ee
This  single expression (\ref{EQ:PoincareRadonAsymptotic}) coincides  with the separate large-$\xi$ asymptotic expansions  (see appendix \ref{SEC:asymptotics}) of $E^k_\nu(\xi)$ when  $k$ is even, and of $O^k_\nu(\xi)$ when  $k$ is odd.  This confirms the truth of the conjecture for odd $k$.

Taking into account the normalization of the Legendre functions  from  appendix \ref{SEC:asymptotics}
we see therefore that the Radon transform takes $\delta$-normalized $L^2[D^2_P]$ eigenfunctions to $\delta$- normalized $L^2[{\rm dS}_2]$ functions times a factor of 
\be
\frac 1{\sqrt {\pi}} \sqrt{\cosh \pi \nu} \Gamma({\textstyle\frac 14+ \frac {i \nu} 2})\Gamma({\textstyle \frac 1 4-\frac {i \nu} 2}).
\ee
This  differs by a factor of $\sqrt{2}$ from the upper half plane  because we are normalizing over is the whole of dS$_2$ rather than half of it.  Once we realize that each $k$ requires only the odd $O^k_\nu$ or only the even $E^k_\nu$ modes, we no longer need the third orthogonality equation  in (\ref{EQ:odd-even-orthogonal}) and can integrate only over $0<\xi<\infty$. In this case the proportionality constants of the upper-half-plane agree with those of the Poincar\'e disc -- as they must do as they are the same transformation, but presented  in different bases.

\subsection{Bound states}

The Schr{\"o}dinger  equations that arise for both the upper-half-plane  and the  hyperbolic polar coordinate transform have bound-state solutions as well as scattering-state solutions.

For the polar coordinates we  can remove the first order derivative in the Legendre equation 
\be
\left(\frac{\partial}{\partial  \sinh \xi} (1+ \sinh^2 \xi) \frac{\partial}{\partial\sinh   \xi}+\frac {k^2} {1+\sinh ^2\xi}\right) \psi(\xi)= -(\nu^2+1/4) \psi (\xi)
\ee
by  writing $\psi(\xi)= (\cosh \xi)^{-1/2} W(\xi)$ so $W$ obeys 
\be
\left(-\frac{d^2W}{d\xi^2} + \frac{1/4-k^2}{\cosh^2 \xi}\right)W(\xi)= \nu^2 W(\xi). 
\ee
This P{\"o}schl–Teller  equation  has scattering solutions $W(\xi)=(\cosh\xi) ^{1/2}P^{k}_{i\nu-\frac 12}(\pm i\sinh \xi)$, or equivalently  $(\cosh\xi) ^{1/2}Q^k_{i\nu-1/2}(\pm i \sinh\xi)$.  

When $k\ge 1$ the potential becomes attractive and we have  bound as well as scattering states. The standard P{\"o}schl–Teller  operator is 
\be
H= -\frac{d^2}{d\xi^2} - \frac{\lambda(\lambda+1)}{\cosh^2 \xi}
\ee
so $\lambda= k-\frac 12$. From the  theory of P{\"o}schl–Teller equations we know  (for example \cite{stone} eq. 19.62) that 
the bound-state eigenvalues are $-\nu_n^2$ ,where $\nu_n$ is given by the $\nu>0$  zeros of
\be
\frac{\Gamma(\nu+1)\Gamma(\nu)}{\Gamma(\nu+\lambda+1)\Gamma(\nu-\lambda)}.
\ee
For $\lambda= k-\frac 12$ this leads to the finite set
$\nu_n = 1/2, 3/2, 5/2,\ldots,  (2k-1)/2$.  

This is an apparently different spectrum than that found for the upper-half plane, which, with the Radon-transform selected self-adjoint extension,  has only half as many bound states: $\nu_n= 3/2, 7/2,\ldots$.   However the  Poincar\'e-patch coordinate 
 system   in the kinematic space of the upper-half plane  covers only {\it half\/}  of AdS$_2$, and the self-adjoint extension in the corresponding eigenvalue problem imposes   boundary condition at infinity --- which corresponds to $\xi=0$  in the present coordinates.
The  P{\"o}schl–Teller equation on the other hand,  sees the whole of AdS$_2$, and hence the denser spectrum. But --- just as in the half-plane case  --- we should  select a subset of the spectrum
that  has even or odd-parity depending on whether $k$ is even or odd. The  lowest energy P{\"o}schl–Teller  bound state  $\nu_k =(2k-1)^2/4$ has even parity,  the next odd, and so on. As a consequence,  the parity  selection  eliminates half of the P{\"o}schl–Teller  bound states and leaves  exactly the same  bound-state spectrum, $\nu_n = 3/2, 7/2,\ldots$, as that found in upper-half plane.

In either realization of hyperbolic space, the Radon transform takes the   complete set of Euclidean-signature Laplace eigenfunctions to only the scattering-state eigenfunctions of the dS$_2$ wave operator. While the  bound states can be obtained as the images non-normalizable modes  such as  $e^{ikx}I_{\nu_n}(|k|y)$ (\cite{jevicki}  Eqn.\ 3.18),  these  are not part of the hyperbolic space mode expansion. The map  
\be
{\mathcal R}:L^2[H^2_+] \to L^2[dS_2]
\ee is therefore not 
surjective.

It is worth noting that the analytic continuation of the continuous-spectrum  singular value  expression
\be
\Lambda(\nu)=\frac 1{\sqrt{2 \pi}} \Gamma\left({\textstyle \frac 14 +\frac{i\nu}{2}}\right)\Gamma\left({\textstyle \frac 14 -\frac{i\nu}{2}}\right)\sqrt{\cosh \pi \nu}
\ee 
has zeros at the the bound-state eigenvalues
$i\nu \to \nu=3/2,7/2$ etc. Taken  literally, when applied to a  bound-state function, the inverse transform  would lead to an infinite multiple of  the  normalized  inverse-image mode. This  is reasonable   as the $e^{ikx}I_{\nu_n}(|k|y)$  are not even $\delta$-normalizable.

\section{Conclusions}

The manner in which in the Poincar\'e-disc Radon transform alternates between two different families, $E^k_\nu(\xi)$ and $O^k_\nu(\xi)$, of special functions depending on whether $k$ is odd or even may seem strange, but in fact it explains exactly what is special about the $\theta=3\pi/4$ parameter choice in \cite{joeP}. The Radon transform does not care about the orientation of the geodesics  when we perform the integrals. The resultant function must therefore be indifferent to which normal to the plane we use in figure \ref{FIG:geodesicmap}. In other words, the transformed function must take the same values at antipodal points in dS$_2$.  When this transformed function is of the form $e^{ik \theta}f(\xi)$, $k\in {\mathbb Z}$, its  values at the pair of antipodal points $(\theta,\xi)$ and $(\theta+\pi,-\xi)$ will be different unless either $k$ is even and $f(\xi)=f(-\xi)$ or $k$ is odd and $f(\xi)= -f(-\xi)$.  The target space of the Radon transform is therefore ${\rm dS}_2/{\mathbb Z}_2$    where ${\mathbb Z}_2$ is the antipodal map. This is not obvious in the spectral decomposition in the upper-half-plane coordinate system, nor in the   $t_1$-$t_2$ dS$_2$ coordinate system in the integral equation in \cite{joeP}. The antipode-identified  space  remains an ${\rm SL}(2,{\mathbb R})$  homogeneous space, which is why it allows the authors of \cite{joeP} the use of the ${\rm SL}(2,{\mathbb R})$  group action to obtain the general eigenfunction from a simple one --- the same strategy we have used in  (\ref{EQ:powerful-trick}) and (\ref{EQ:powerfultrick2}). 

The  antipodal  identification  arises also in the non-local mapping of general-dimensional  Euclidean de Sitter space to Lorentzian-signature de Sitter, as is  discussed extensively in \cite{antipode}. It would be interesting to relate those ideas the space-time duality aspects of SYK.

\section{Acknowledgements} 

This work was not directly supported by any external funding agency, but it would not have been possible without resources provided by the Department of Physics at the University of Illinois at Urbana-Champaign.  I would like to thank  Marc Klinger, Gabriele La Nave, Yuting Bai and Sami Muslih for valuable conversations.

\appendix

\section{Group Decompositions and Cosets}
\label{SEC:decompositions}

 Coordinate systems for  ${\rm SL}(2,{\mathbb R})\simeq {\rm SU}(1,1)$ and their cosets  may be   obtained  from various group decompositions involving the one-parameter subgroups. The corresponding left- and right-invariant vector fields  combine to give the     Casimir differential operators whose eigenfunctions provide  representation matrix elements which are the familiar families of special-functions\cite{vilenkin}. Here we describe the decompositions used in the main text, and some other closely related versions.

\subsection{Subgroups}

\subsubsection{One-parameter subgroups of ${\rm SL}(2, {\mathbb R})$}

Some standard named subgroups are 
\bea
N:&&
n(t)= \exp\{-t L_1\}=\left(\begin{matrix}1&0\cr t &1\end{matrix}\right),\nonumber\\
\tilde N:&&  n^T(t)  =\exp\{t L_{-1}\}=\left(\begin{matrix}1&t\cr 0 &1\end{matrix}\right),
\nonumber\\
H:&&  h(s) = \exp \{sL_0\}= \left(\begin{matrix} e^{s/2} &0\cr 0& e^{-s/2}\end{matrix}\right), 
 \nonumber\\
 A:&& a(u) = \exp\{ u(L_{-1}- L_1)/2\}= \left(\begin{matrix} \cosh u/2 & \sinh u/2\cr \sinh u/2 &\cosh u/2\end{matrix}\right),
 \nonumber\\
 T:&& t(y) = \left(\begin{matrix}y^{1/2}&0\cr 0& y^{-1/2}\end{matrix}\right), \quad y>0,\nonumber\\
 K:&&  k(\theta) =  \exp\{\theta (L_1+L_{-1})\}= \left(\begin{matrix} \cos \theta/2& \sin \theta/2 \cr -\sin\theta/2 &\cos\theta/2\end{matrix}\right).\nonumber  
\eea

 \subsubsection{ One parameter subgroups of ${\rm SU}(1,1)$}
 \bea
 \exp\{\phi \Lambda_0\}&=&  \left(\begin{matrix} e^{i\phi/2} & 0\cr 0&e^{-i\phi/2}\end{matrix}\right),\nonumber\\
 \exp\{\theta \Lambda_2\}&=&  \left(\begin{matrix} \cosh(\theta/2) & i \sinh (\theta/2)\cr 
 - i\sinh(\theta/2) &\cosh(\theta/2)\end{matrix}\right),\nonumber\\
 \exp\{\xi \Lambda_1\}&=& \left(\begin{matrix} \cosh(\xi/2) &  \sinh (\xi/2)\cr 
 \sinh(\xi/2) &\cosh(\xi/2)\end{matrix}\right). \nonumber
\eea
 The first  is an elliptic subgroup and  the other  two are hyperbolic.

 \subsection{Specific decompositions}


\subsubsection{Euler-angle decomposition}
\label{SUBSUBSEC:euler}

This is a decomposition of ${\rm SU}(1,1)$ that provides the eigenvalue problem on the Poincar{\'e} disc. We factorise
\be
 g(\phi,\xi,\theta)= \exp\{ \phi \Lambda_0\} \exp\{ \xi \Lambda_1\}\exp\{ -\theta \Lambda_0\} =
 \left(\begin{matrix} e^{i(\phi-\theta)/2}\cosh \xi/2 & e^{i(\phi+\theta)/2}\sinh \xi/2\cr e^{-i(\phi+\theta)/2} \sinh\xi/2 & e^{-i(\phi-\theta)/2}\cosh\xi/2\end{matrix}\right).
 \ee
 The point $z=0$ is left fixed by $\exp\{ -\theta \Lambda_0\}\in {\rm U}(1)$ and its orbit is the set $e^{i\phi}\tanh \xi/2$ which are coordinates on the Poincar{\'e} disc.   

The group metric in these coordinates is 
\bea
ds^2&=&2\,{\rm tr}\{(g^{-1} dg)( g^{-1} dg)\}\nonumber\\
&=&- d\theta^2+d\xi^2-d\phi^2+2\cosh \xi d\theta d\phi.
\eea
The factor of two is inserted to compensate for the factor of ${\textstyle \frac 12}$ in   ${\rm tr}\{ \Lambda_i \Lambda_2\}= {\textstyle \frac 12}  \eta_{ij}$ with $\eta= {\rm diag}(-1,1,1)$.

 The Poincare disc  is the coset ${\rm SU}(1,1)/U(1)$. The coset metric  is found by  setting  $d\theta = d\phi \cosh  \xi$ because  this   minimizes $ds^2$ at fixed $d\xi$ and $d\phi$.
We  find 
\be
ds^2= d\xi^2+\sinh^2 \xi d\phi^2   
\ee
which is the usual hyperbolic polar coordinate metric.

The left- and right-invariant  vector fields are 
\bea
\lambda_0 &=& - \partial_\theta \nonumber\\
\lambda_1 &=&-\coth {\xi}\sin \theta  \partial_\phi+\cos\theta \partial_\xi-\coth \xi \sin\theta  \partial_\theta\nonumber\\
\lambda_2 &=& \cos\theta \csch\xi \partial_\phi +\sin\theta \partial_\xi +\cos\theta \coth \xi \partial_\theta
\eea
and
\bea 
\rho_0 &=&  \partial_\phi \nonumber\\
\rho_1 &=&-\coth \xi \sin \phi\partial_\phi+\cos\phi \partial_\xi-\csch\xi \sin\phi  \partial_\theta\nonumber\\
\rho_2 &=& \cos\phi  \coth\xi \partial_\phi +\sin\phi \partial_\xi +\cos\phi \csch \xi  \partial_\theta 
\eea
The Casimir is 
\be
C_2= \Lambda_0^2 - \Lambda_1^2-\Lambda^2_2
\mapsto \lambda_0^2- \lambda_1^2-\lambda_2^2
=   \rho_0 ^2-\rho_1^2- \rho_2^2.
\ee
Acting on functions of the form $e^{i\lambda \phi} \psi(\xi) e^{i\mu\theta}$ the Casimir becomes   
\be
C_2 \to -\frac 1{\sinh \xi}\frac{\partial}{\partial  \xi} \sinh \xi  \frac{\partial}{\partial  \xi}+\frac 1{\sinh^2\xi} (\lambda^2+\mu^2 +2\lambda\mu \cosh \xi).
\ee
Functions on the coset ${\rm SU}(1,1)/U(1)$ have  $\mu=0$. Non zero values of $\mu$ would correspond to sections of a line bundle over the coset space.

\subsubsection{AdS$_2$ via ${\rm SU}(1,1)$ }
\label{SUBSEC:AdSviaSU11}

AdS$_2$ is a quotient of  ${\rm SU}(1,1)$ by a hyperbolic subgroup. For example   $\exp\{ -\tau \Lambda_2\}$  is hyperbolic because it fixes the points $\pm i$ on the  boundary of the Poincar{\'e} disc. If we factorize  
\be
g(t,\xi,\tau)= \exp\{ t \Lambda_0\} \exp\{ \xi \Lambda_1\}\exp\{ -\tau \Lambda_2\}
\ee
we  find the group metric $2 {\rm tr}\{ (g^{-1}dg)^2\}$ to be
\be
ds^2= d\tau^2+d\xi^2 -dt^2 - 2\sinh \xi d\tau dt. 
\ee
The metric on the coset which forgets $\tau$ is 
\be
ds^2=d\xi^2- \cosh^2\xi dt^2,
\ee
which   is a global AdS$_2$ metric. 
If we rename $t\to \theta$ and define an angle $\alpha$ by $\csch \xi= \tan \alpha$ the this metric becomes 
\be
ds^2={\rm csc}^2 \alpha(-d\theta^2+ d\alpha^2)
\label{EQ:alpha-theta-metric2}
\ee
which is the metric on the space of geodesics on the Poincar\'e disc parameterized as in figure \ref{FIG:alpha-theta}.

The left-invariant vector fields are 
\bea
\lambda_0&=&\cosh \tau \,\sech \xi \partial_t -\sinh \tau \partial_\xi + \cosh \tau \,\tanh \xi \partial _\tau \nonumber\\
\lambda_1&=&- \sech \xi\, \sinh\tau \partial_t+ \cosh \tau \partial_\xi- \sinh \tau \,\tanh\xi \partial _\tau\nonumber\\
\lambda_2&=&- \partial_\tau
\eea
The right-invariant vector fields are 
\bea
\rho_0&=&\partial_t \nonumber\\
\rho_1&=&- \sin t\, \tanh \xi \partial_t +\cos t \partial_\xi +\sech \xi \partial_\tau\nonumber\\
\rho_2&=&\cos t \,\tanh\xi \partial_t +\sin t \partial_\xi -\cos t \,\sech \xi \partial_\tau
\eea
From either the left or right invariant fields we find that the Casimir acting on 
 functions of the form $e^{i\lambda t} \psi(\xi) e^{i\mu\theta}$ is
\be
C_2 \to -\frac 1{\cosh  \xi}\frac{\partial}{\partial  \xi} \cosh \xi  \frac{\partial}{\partial  \xi}-\frac 1{\cosh^2\xi} (\lambda^2-\mu^2 +2\lambda\mu \sinh \xi) .
\ee
When $\mu=0$ we have functions  on the coset and the eigenvalue equation becomes  
\be
-\left(\frac{\partial}{\partial  \sinh \xi} (1+ \sinh^2 \xi) \frac{\partial}{\partial\sinh   \xi}+\frac {\lambda^2} {1+\sinh ^2\xi}\right) \psi= (\nu^2+1/4)\psi 
\ee
with solutions $\psi=P^\lambda_{i\nu-1/2}(\pm i \sinh \xi)$ or $Q^\lambda_{\i\nu-1/2}(\pm i \sinh \xi)$.   

\subsubsection{AdS$_2$ via ${\rm SL}(2,{\mathbb R})$}

The subgroup $h(t)$ is hyperbolic and we can parameterize as  
\be
g(t,\rho,\phi)= h(t)a(\rho) h(\phi) =\left(\begin{matrix} e^{(t+\phi)/2} \cosh(\rho/2) & e^{(t-\phi)/2} \sinh(\rho/2)\cr
 e^{(\phi-t)/2} \sinh(\rho/2)& e^{-(t+\phi)/2} \cosh(\rho/2)\end{matrix}\right)
 \ee
 This is not a global parameterization  because the lower right entry in $g$ is always positive.
 The group metric  becomes  
 \be
 ds= dt^2+d\rho^2 +d\phi^2 +2 dt d\phi \cosh \theta,
 \ee
  and  the coset  ${\rm SL}(2,{\mathbb R})/H$ has metric
  \be
  ds^2 =d\rho^2 - \sinh^2(\rho) dt^2,
  \ee
  which  the metric of a  Rindler patch of AdS$_2$.
   
 \subsubsection{Iwasawa decomposition} 
 The 
 Iwasawa NAK decomposition is 
 \be
 g(x,y,\theta)= n^T(x)t(y)k(\theta)=\frac  1{\sqrt y}\left(
\begin{matrix}
 y \cos\theta/2-x \sin\theta/2& x \cos
  \theta/2 +y \sin\theta/2\cr
 -\sin\theta/2 & \cos\theta/2\end{matrix}
\right)= \left(\begin{matrix}a&b\cr c&d\end{matrix}\right)
\ee

If we let this act as 
\be
z\to \frac {az+b}{cz+d}
\ee
on the point $z=i$ we get
\be
\frac{i(y \cos\theta/2-x \sin\theta/2) + x \cos
  \theta/2 +y \sin\theta/2}{
 -i\sin\theta/2 + \cos\theta/2} = x+iy,
\ee
which is independent of $\theta$. As $y>0$ the orbit of $z=i$ is the upper half-plane revealed as the coset $H_+^2= {\rm SL}(2,{\mathbb R})/K$. 
The metric on the group is 
 \be
 ds^2 =2 \,{\rm \tr}\{g^{-1}dg\, g^{-1}dg\}=\frac{dy^2- 2 y dx d\theta -y^2 d\theta^2}{y^2}.
 \ee
 To get the  metric on the coset  we set  $d\theta = -dx/y$  as this  choice 
 minimizes the distance $d^2s$ at fixed $dx$, $dy$.
 Then 
 \be
 \frac{dy^2- 2 y dx d\theta -y^2 d\theta^2}{y^2}\to\frac{dx^2+dy^2}{y^2},
 \ee
 which is the upper-half-plane metric.
 
 The left invariant vector fields are 
 \bea
 {\mathcal L}_{-1}&=& y \cos \theta \, \partial_x + y \sin\theta  \,\partial_yu 
 + 2 \sin^2\theta/2 \,  \partial_\theta \nonumber\\
  {\mathcal L}_{0}&=& -y \sin\theta \, \partial_x +  y\cos\theta \, \partial_\theta \nonumber\\
  {\mathcal L}_{1}&=& -y \cos
  \theta\,  \partial_x-y\sin\theta \partial_y + 2 \cos^2 \theta/2 \,\partial_\theta 
  \eea
 The right invariant fields are 
 \bea
 {\mathcal R}_{-1}&=&  \partial_x  \nonumber\\
  {\mathcal R}_{0}&=& x  \, \partial_x +  y\, \partial_y \nonumber\\
  {\mathcal R}_{1}&=& (x^2-y^2)\,  \partial_x+2yx\,  \partial_y + 2 y \,\partial_\theta 
  \eea
 The quadratic Casimir operator is
 \bea
 C_2&\to &-{\mathcal L}^2_0+ \frac 12( {\mathcal L}_1 {\mathcal L}_{-1}+  {\mathcal L}_{-1} {\mathcal L}_{1})\mapsto 
 -{\mathcal R}^2_0+ \frac 12( {\mathcal R}_1 {\mathcal R}_{-1}+  {\mathcal R}_{-1} {\mathcal R}_{1})\nonumber\\
 &=&- y^2\left(\frac{\partial^2}{\partial x^2} + \frac{\partial^2}{\partial y^2}\right)-2y\frac{ \partial^2}{\partial x\partial \theta}.
 \eea 
Functions on the ${\rm SL}(2,{\mathbb R})/K $ coset are independent of $\theta$. Allowing $\theta$-dependence $\propto e^{ik\theta}$ gives us the Schr\"odinger equation  of a charged particle in a   background magnetic field \cite{comtet1,comtet2,comtet3}. To mathematicians the resulting wavefunctions  are Maa{\ss} waveforms of level-$k$ \cite{fay}.
 
  \subsubsection{NHK decomposition} 
  \label{SUBSUBSEC:NHK}
 \be
 g(t,\theta,\phi)= n^T(t) h(\xi) k(\phi)= \left(\begin{matrix} e^{\xi/2} \cos(\phi/2)-t e^{-\xi/2} \sin (\phi/2)& e^{\xi/2} \sin(\phi/2)+
 t e^{-\xi/2} \cos(\phi/2) \cr
e^{\theta/2} t\cos(\phi/2)  & e^{-\xi/2} \cos(\phi/2)\end{matrix} \right)
  \ee
  The group metric is 
 \be
 ds^2 = d\xi^2 -d\phi^2  - 2d\phi dt e^{-\xi},
 \ee
  The metric on $SL(2,{\mathbb R})/K$
  \be
  ds^2= d\xi^2 + e^{-2\xi} dt^2 
  \ee
 The Casimir  acting on $\psi= e^{i\lambda t}f(\theta) e^{i\mu\phi}$
 is 
 \be
 C_2\to \left(-\frac{d^2 }{d\theta^2} + \frac{d f}{d\theta}  -2\lambda \mu e^{\theta}+\lambda^2 e^{2\theta}\right)f
 \ee
  
   \subsubsection{NHA decomposition}
   \label{SUBSUBSEC:NHA} 
 
 \be
 g= n^T(t) h(\xi) a(\phi)= \left(\begin{matrix} e^{\xi/2} \cosh(\phi/2)+t e^{-\xi/2} \sinh(\phi/2),  & e^{\xi/2} \sinh(\phi/2)+t e^{-\xi/2} \cosh (\phi/2)  \cr
 e^{-\xi/2} \sinh (\phi/2) & e^{-\xi/2} \cosh(\phi/2) \end{matrix}\right)
  \ee
 The factor  $a(\phi)\in A$ leaves the points $x=\pm 1, y=0$ fixed and the  coset $SL(2,{\mathbb R})/A$ parametrises the geodesics on $H^2_+$. 
  
 The group metric is
 \be
 ds^2 = d\xi^2 +d\phi^2  + 2d\phi \,dt e^{-\xi}.
 \ee
  The coset metric on $SL(2,{\mathbb R})/A$ is found by setting $d\phi\to - e^{-\xi}dt$ so  
  \be 
  ds^2= d\theta^2 - e^{-2\xi} dt^2 
  \ee
  
  The left invariant vector fields are 
  \bea
  {\mathcal L}_{-1}&=& e^\xi\cosh \phi \,\partial_t -\sinh \phi \,\partial_\xi - (1- \cosh \phi) \partial_\phi,\nonumber\\
  {\mathcal L}_{0}&=& - e^\xi \sinh \phi  \, \partial_t +  \cosh \phi\, \partial_\xi +\sinh \phi \,\partial \phi, \nonumber\\
  {\mathcal L}_{1}&=& e^\xi \cosh \phi \, \partial_t-\sinh \phi \,\partial_\xi-(1+\cosh  \phi )\partial_\phi  .
\eea
  The right invariant vector fields are 
  \bea
  {\mathcal R}_{-1}&=& \partial_t, \nonumber\\
  {\mathcal R}_{0}&=& t \,  \partial_t +  \partial_\xi,  \nonumber\\
  {\mathcal R}_{1}&=& e^{2\xi} \partial_t+2t \, \partial_\xi- 2e^\xi \partial_\phi  .
\eea

  The Casimir  acting on $\psi= e^{i\lambda t}f(\theta) e^{i\mu\phi}$
 is 
 \be 
C-2\to - \frac{d^2 }{d\theta^2} + \frac{d }{d\theta} +2\lambda \mu e^{\theta}-\lambda^2 e^{2\theta}
 \ee
Note the sign change in the $e^{+2\theta}$ potential as we pass from Euclidean AdS to Lorentz AdS.

\section{Asymptotics and orthogonality of the $E^k_\nu(\xi)$ and $O^k_\nu(\xi)$
 functions}
\label{SEC:asymptotics}

The $\xi\gg1 $ behaviours of the $E^k_\nu(\xi)$ and $O^k_\nu(\xi)$ functions can be found from  \cite{chukhrukidze}   after one observes  that in the large-$\xi$  limit the $_2F_1$ hypergeometric functions in that paper  can be replaced by unity\footnote{Unfortunately \cite{chukhrukidze} eq.\ $13$ does not   agree with  numerical evaluations. Those presented in  our eqs. B1-4 involve  a  slight modification to the choice of  branches of $i^{i\nu-1/2}$ in eq.\ 13   so as   to agree with the numerical evaluations in both phase and amplitude.}. They are different for odd and even $k$. 

For even integer $k$ we have 
 \bea 
 E^k_\nu(\xi) &\stackrel{\rm def}{=}& 
\frac  1 2\left\{P^{k}_{i\nu-{\textstyle \frac 12}}(i \sinh \xi)+ P^{k}_{i\nu-{\textstyle \frac 12}}(-i \sinh \xi)\right\}\nonumber\\
 &\sim& \frac{(-1)^{k/2}}{\sqrt{2\pi \cosh \xi}} \left\{ e^{i\nu \xi} \sin\left( \pi ({\textstyle \frac 14 +\frac{i\nu}2})\right) \frac {\Gamma (i\nu)}{\Gamma({\textstyle \frac 12} -k +i\nu)} + e^{-i\nu \xi} \sin\left( \pi ({\textstyle \frac 14 -\frac{i\nu}2})\right) \frac {\Gamma (-i\nu)}{\Gamma({\textstyle \frac 12} -k -i\nu)} \right\},\nonumber\\
 \label{EQ:B1}
 \eea 
and 
 \bea 
 O^k_\nu(\xi) &\stackrel{\rm def}{=}& 
 \frac {1 }  {2i}\left\{P^{k}_{i\nu-{\textstyle \frac 12}}(i \sinh \xi)- P^{k}_{i\nu-{\textstyle \frac 12}}(-i \sinh \xi)\right\}\nonumber\\
 &\sim& \frac{(-1)^{k/2}}{\sqrt{2\pi \cosh \xi}} \left\{ e^{i\nu \xi} \sin\left( \pi ({\textstyle \frac 14 -\frac{i\nu}2})\right) \frac {\Gamma (i\nu)}{\Gamma({\textstyle \frac 12} -k +i\nu)} + e^{-i\nu \xi} \sin\left( \pi ({\textstyle \frac 14 +\frac{i\nu}2})\right) \frac {\Gamma (-i\nu)}{\Gamma({\textstyle \frac 12} -k -i\nu)} \right\}.\nonumber\\
 \label{EQ:B2}
 \eea 
 
 For odd integer $k$ we have
 \bea 
 E^k_\nu(\xi) &\stackrel{\rm def}{=}&
\frac {1 } 2\left\{P^{k}_{i\nu-{\textstyle \frac 12}}(i \sinh \xi)+ P^{k}_{i\nu-{\textstyle \frac 12}}(-i \sinh \xi)\right\}\nonumber\\
 &\sim& \frac{(-1)^{(k+1)/2}}{\sqrt{2\pi \cosh \xi}} \left\{ e^{i\nu \xi} \sin\left( \pi ({\textstyle \frac 14 -\frac{i\nu}2})\right) \frac {\Gamma (i\nu)}{\Gamma({\textstyle \frac 12} -k +i\nu)} + e^{-i\nu \xi} \sin\left( \pi ({\textstyle \frac 14 +\frac{i\nu}2})\right) \frac {\Gamma (-i\nu)}{\Gamma({\textstyle \frac 12} -k -i\nu)} \right\},\nonumber\\
\label{EQ:B3}
 \eea 
 and 
 \bea 
 O^k_\nu(\xi) &\stackrel{\rm def}{=}& 
 \frac {1} {2i}\left\{P^{k}_{i\nu-{\textstyle \frac 12}}(i \sinh \xi)- P^{k}_{i\nu-{\textstyle \frac 12}}(-i \sinh \xi)\right\}\nonumber\\
 &\sim& \frac{(-1)^{(k+1)/2}}{\sqrt{2\pi \cosh \xi}} \left\{ e^{i\nu \xi} \sin\left( \pi ({\textstyle \frac 14 +\frac{i\nu}2})\right) \frac {\Gamma (i\nu)}{\Gamma({\textstyle \frac 12} -k +i\nu)} + e^{-i\nu \xi} \sin\left( \pi ({\textstyle \frac 14 -\frac{i\nu}2})\right) \frac {\Gamma (-i\nu)}{\Gamma({\textstyle \frac 12} -k -i\nu)} \right\}.\nonumber\\
 \label{EQ:B4}
 \eea

An integration by parts shows that
 \be
 (\mu^2-\nu^2)\int_{-\infty}^{\infty} E^k_{\mu}(\xi)E^k_\nu(\xi) \cosh \xi d\xi =
 \left[\cosh \xi(E^k_\mu \partial_\xi E_\nu^k- E^k_\nu \partial_\xi E_\mu^k)\right]_{-\infty}^{\infty},
 \ee
 and similarly for the $O^k_\nu$.  Substituting in the asymptotic expressions and using
 \be
 \lim_{\xi\to \infty} \frac {2\sin\left( \xi (\mu -\nu)\right)}{(\mu-\nu)}= 2\pi \delta(\mu-\nu)
 \ee
 we find  that the orthogonality relations are 
\bea
\int_{-\infty}^{\infty} E^k_\mu(\xi)E^k_{\nu}(\xi)\cosh \xi d\xi &=& \frac{\pi}{\nu\, \tanh\pi \nu} \frac {\delta(\mu-\nu)}{\Gamma({\textstyle \frac 12}+i \nu-k)\Gamma({\textstyle \frac 12}-i \nu-k)},\nonumber\\
\int_{-\infty}^{\infty} O^k_\mu(\xi)O^k_{\nu}(\xi)\cosh \xi d\xi &=& \frac{\pi}{\nu\, \tanh\pi \nu} \frac {\delta(\mu-\nu)}{\Gamma({\textstyle \frac 12}+i \nu-k)\Gamma({\textstyle \frac 12}-i \nu-k)},\nonumber\\
\int_{-\infty}^{\infty} E^k_\mu(\xi)O^k_{\nu}(\xi)\cosh \xi d\xi&=&0,
\label{EQ:odd-even-orthogonal}
\eea
where we have assumed that both $\mu$ and $\nu$ are positive.
The extra factor of $ \cosh \pi \nu$ compared to the  $P^k_{i\nu-1/2}(\cosh \xi)$ normalization comes from
\be
| \sin\left( \pi ({\textstyle \frac 14 -\frac{i\nu}2})\right)|^2=\frac 12 \cosh \pi \nu.
\ee
and the fact that there are contributions from both $\pm \infty$.


\begin{thebibliography}{99}

\bibitem{kitaev-syk} A. Kitaev, Seminars {\it A simple model of quantum holography\/}. Seminars  in the  KITP {\it Strings  and Entanglement\/} program 2015.  (Feb. 12, April 7, and May 27, 2015)  
\url{http://online.kitp.ucsb.edu/online/entangled15/}


\bibitem{sy} S. Sachdev, J. Ye, {\it Gapless spin-fluid ground state in a random quantum Heisenberg magnet\/}, Phys.\ Rev.\ Lett.\ 70 (1993) 3339–3342. 

\bibitem{maldacena} J.~Maldacena, {\it Remarks on the Sachdev-Ye-Kitaev model\/}, 
 Phys.\ Rev.\ D 94 (2016) 10, 106002;
ArXiv:1604.07818.

\bibitem{rosenhaus} V.~Rosenhaus, {\it
An introduction to the SYK model},  J. Phys.\ A: Math.\ Theor.\ 52 (2019) 323001.

\bibitem{sarosi} G.~S{\'a}rosi, {\it AdS$_2$ holography and the SYK model\/}, ArXiv:1711.08482 

\bibitem{joeP}  J.~Polchinski, V.~Rosenhaus, {\it The Spectrum in the Sachdev-Ye-Kitaev Model\/},  
JHEP04(2016)001; 
 ArXiv 1601.06768
 
 \bibitem{dunster90} T.~M.~Dunster, {\it Bessel functions of purely imaginary order, with an application to second- order linear differential equations having a large parameter\/}, SIAM J. Math. Anal., 21 (1990), 995-1018.
 
 \bibitem{gitman} D.~M.~Gitman, I.~V.~Tyutin, B.~L.~Voronov, {\it Self-adjoint Extensions in Quantum Mechanics\/},  (Birkhäuser, Basel, 2012). 
 
 \bibitem{andrianov} A.~A.~Andrianov, C.~Lan, O.~O.~Novikov, Y-F.~Wang, {\it Integrable minisuperspace models with Liouville field: energy density self-adjointness and semiclassical wave packets\/}, Eur.\ Phys.\ J.\ C 78 (2018) 786.

 \bibitem{kobayashi} H.~ Kobayashi, I.~Tsutsui, {\it Quantum mechanical Liouville model with attractive potential\/}, Nuclear Physics B {\bf 472} (1996) 409-426; arXiv:hep-th/9601111
 
 \bibitem{jevicki} S.~R.~Das, A.~Ghosh, A.~Jevicki, K. Suzuki, {\it Space-Time in the SYK Model}, JHEP07(2018)184;
 arXiv:1712.02725.

\bibitem{helgason} S.~Helgason, {\it The Radon Transform}, 2nd edition, Springer (NY) 2013. 

\bibitem{gelfand}  I.~M.~Gel'fand,
M.~I.~Graev,
N.~Ya.~Vilenkin, {\it 
Generalized Functions, Volume 5: Integral Geometry and Representation Theory}, AMS Chelsea Publishing (1966).


\bibitem{hyperbolic_radon} E.~C.~Tarabusi,  M.~A.~Picardello, {\it Radon Transforms in Hyperbolic Spaces and Their Discrete Counterparts\/}, Complex Analysis and Operator Theory (2021) 15:13

\bibitem{watson} G.~N.~Watson, {\it A Treatise on the Theory of Bessel Functions}, Cambridge University Press 1922.




\bibitem{kitaev} A.~Kitaev,  
{\it Notes on $\widetilde{SL(2, R)}$ representations}, arXiv:1711.08169

\bibitem{integral-geometry} B.~Czech, L.~Lamprou, S.~McCandlish, J.~Sully, {\it Integral geometry and holography\/}, 
JHEP10(2015)175;
ArXiv 1505.05515.

\bibitem{bargmann} V. Bargmann, {\it Irreducible Unitary Representations of the Lorentz Group\/},
Annals of Mathematics
(Second Series) {\bf 48} (1947) 568-640.



\bibitem{hyperbolic_radon_AdS} B.~Czech, L.~Lamprou, S.~McCandlish, B.~Mosk,
J.~Sully, {\it A Stereoscopic Look into the Bulk\/}, JHEP07(2016)129; arXiv:1604.03110.


\bibitem{Ma1} X.~Huang, C-T.~Ma, {\it Berry Curvature and Riemann Curvature in Kinematic Space with Spherical Entangling Surface\/},  Progress of Physics, 69 (2021)
2000048; arXiv:2003.12252

\bibitem{Ma2} X.~Huang, C-T.~Ma,  {\it The probe of curvature in the Lorentzian AdS2/CFT1 correspondence\/},
 Phys.\ Lett.\  B
(2019) 134936; arXiv:1907.01422




\bibitem{dunster2013} T.~M.~Dunster, 
{\it Conical functions of purely imaginary order and argument\/}, Proceedings of the Royal Society of Edinburgh, 143A (2013)
929-955.

\bibitem{NIST}  {\it Digital Library of Mathematical Functions\/}, \url{https://dlmf.nist.gov}

\bibitem{oberhettinger}  F.~Oberhettinger, T.~P.~Higgins, {\it Tables of Lebedev, Mehler, and Generalized Mehler Transforms\/}, Boeing  Scientific Research report D1-82-0136 (1961).
\url{https://apps.dtic.mil/sti/tr/pdf/AD0267210.pdf}

\bibitem{antipode} V.~Balasubramanian, J.~de Boer, D.~Minic, {\it Exploring de Sitter Space and Holography\/}, Annals of Physics 303, (2003) 59-116; arXiv:0207245

\bibitem{stone}  
P.~Goldbart, M.~Stone, {\it Mathematics for Physics\/}, Cambridge University Press ( 2009)





\bibitem{vilenkin} N.~Ya.~Vilenkin, A.~U.~Klimyk, {\it Representation of Lie Groups and Special Functions\/}, Kluwer Academic, London,
1991 Vols. I, II, III.

\bibitem{comtet1} A.~Comtet, {\it   On the Landau Levels on the Hyperbolic Plane\/},  Annals of Physics (NY)  173 (1987) 185-209.

\bibitem{comtet2} A.~Comtet, 
{\it Effective action on the hyperbolic plane in a constant external field\/}, J. Math.\ Phys.\ 26  (1985) 185-191.

\bibitem{comtet3} A.~Comtet, {\it Effective action on the hyperbolic plane in a constant external field\/}, J.\ Math.\ Phys.\ 26, (1985)185-191.

\bibitem{fay} J.~D.~Fay, {\it  Fourier coefficients of the resolvent for a Fuchsian group\/}, Journal f\"ur die reine und angewandte Mathematik 
 {\bf 0293\_0294} (1977) 143-203.
 


\bibitem{chukhrukidze} N. K. Chukhrukidze, {\it Asymptotic expansions of Legendre Spherical Functions with an imaginary Argument\/}, USSR Computational Mathematics and Mathematical Physics, 8 (1968)1-13.










\end{thebibliography}
\end{document}